\documentclass[pra,showpacs,twocolumn,superscriptaddress,nofootinbib]{revtex4}

\usepackage{amsmath,amsfonts,amssymb,amstext}
\usepackage{graphicx}
\usepackage[colorlinks,linkcolor=blue]{hyperref}
\usepackage{multirow}
\usepackage{lmodern}

\newcommand{\ket}[1]{\left|#1\right\rangle}
\newcommand{\bra}[1]{\left\langle#1\right|}
\newcommand{\braket}[2]{\left\langle#1\right|\left.#2\right\rangle}
\newcommand{\ketbra}[2]{\left|#1\right\rangle\!\left\langle#2\right|}

\newcommand{\avg}[1]{\left\langle #1 \right\rangle}

\def\cO{{\cal O}}
\DeclareMathOperator{\Tr}{Tr}

\def\({\left(}
\def\){\right)}

\renewcommand{\vec}[1]{\mathbf{#1}}

\newcommand{\be}{\begin{equation}}
\newcommand{\ee}{\end{equation}}
\newcommand{\bea}{\begin{eqnarray}}
\newcommand{\eea}{\end{eqnarray}}
\newcommand{\bF}{\begin{figure}}
\newcommand{\eF}{\end{figure}}
\newcommand{\dg}{\dagger}
\newcommand{\bi}{\begin{itemize}}
\newcommand{\ei}{\end{itemize}}

\newcommand{\mbf}[1]{\mathbf{#1}}

\begin{document}
\title{Quantum tomography of the full hyperfine manifold of atomic spins via continuous measurement on an ensemble}
\date{\today}

\author{Carlos A. Riofr\'{i}o}
\affiliation{Center for Quantum Information and Control (CQuIC) and Department of Physics and Astronomy, University of New Mexico, Albuquerque, NM, 87131, USA}

\author{Poul S. Jessen}
\affiliation{Center for Quantum Information and Control (CQuIC) and College of Optical Sciences, University of Arizona, Tucson, AZ, 85721, USA}

\author{Ivan H. Deutsch}
\affiliation{Center for Quantum Information and Control (CQuIC) and Department of Physics and Astronomy, University of New Mexico, Albuquerque, NM, 87131, USA}

\begin{abstract}
Quantum state reconstruction based on weak continuous measurement has the advantage of being fast, accurate, and almost non-perturbative.  In this work we present a pedagogical review of the protocol proposed by Silberfarb {\em et al.}, PRL {\bf 95} 030402 (2005), whereby an ensemble of identically prepared systems is collectively probed and controlled in a time-dependent manner so as to create an informationally complete continuous measurement record.  The measurement history is then inverted to determine the state at the initial time through a maximum-likelihood estimate.  The general formalism is applied to the case of reconstruction of the quantum state encoded in the magnetic sublevels of a large-spin alkali atom, ${}^{133}$Cs.  We detail two different protocols for control.  Using magnetic interactions and a quadratic ac-Stark shift, we can reconstruct a chosen hyperfine manifold  $F$, e.g., the 7-dimensional $F=3$ manifold in the electronic-ground state of Cs.  We review the procedure as implemented in experiments  (Smith {\em et al.}, PRL {\bf 97} 180403 (2006)). We extend the protocol to the more ambitious case of reconstruction of states in the full 16-dimensional electronic-ground subspace ($F=3 \oplus F=4$), controlled by microwaves and radio-frequency magnetic fields. We give detailed derivations of all physical interactions, approximations, numerical methods, and fitting procedures, tailored to the realistic experimental setting.  For the case of light-shift and magnetic control, reconstruction fidelities of $\sim 0.95$ have been achieved, limited primarily by inhomogeneities in the light shift.  For the case of microwave/RF-control we simulate fidelity $>0.97$, limited primarily by signal-to-noise.
\end{abstract}

\maketitle

\section{Introduction}

An essential tool in quantum information science is quantum tomography (QT) \cite{paris04}.  The ability to estimate a quantum state  is required to diagnose quantum information processors and evaluate the fidelity of a given protocol.  The fundamental information-gain/disturbance tradeoff in a measurement of a quantum system implies that any QT protocol requires multiple, nearly identical copies of the state.   Typically, this procedure is carried out through a series of strong destructive measurements of an informationally complete set observables acting on repeatedly prepared copies of the system.  For this reason, QT is generally a time consuming and tedious procedure when applied to large dimensional systems \cite{haffner05} and even more so when extended to quantum-process tomography in which a whole collection of quantum states must be analyzed \cite{dariano04}.  

In some platforms, one has the ability to probe a large ensemble of identical systems simultaneously.  In this case, one can enhance the collection of statistics, e.g., for estimation of the probability of occupation in an eigenstate under projective measurements.  While such projective measurements on ensembles can be used for QT \cite{klose01}, in principle one can dramatically improve the speed, robustness, and experimental complexity of QT by instead employing weak-continuous measurement.  In a protocol originally developed by Silberfarb {\em et al.} \cite{silberfarb05}, an atomic ensemble undergoes a chosen dynamical evolution to generate an informationally complete measurement record.  An algorithm is then used to invert the measurement history to determine the maximum likelihood of the initial state.  For moderately large ensembles, one can attain a sufficient signal-to-noise ratio to enable extraction of the required information, while simultaneously maintaining the quantum projection noise below the intrinsic noise of the quantum probe.  In this case, quantum backaction is negligible and in principle, one can extract the necessary data for QT in a single run of the experiment on a single ensemble.  

We have previously employed a continuous measurement protocol to perform QT on the 7-dimensional, $F=3$ atomic hyperfine spin manifold, in an ensemble of cesium atoms \cite{silberfarb05}.  With this tool in hand, we diagnosed the performance of state-to-state quantum maps, designed and implemented by optimal control techniques \cite{smith06}.  More recently,  this QT protocol was a central component that enabled us to measure the time evolution of the quantum state of the spin undergoing the quantum chaotic dynamic of a nonlinear kicked top \cite{chaudhury09}.  Observing dynamics of a density matrix for any reasonable duration would have been formidable without an efficient method for QT at each time step.  

Our goal here is to give a detailed and pedagogical description of the protocol, focusing on the ingredients that are necessary to enable QT of complex systems, and its application to atomic spin systems and other platforms. The remainder of this paper is organized as follows. In Sec. II,  we give a detailed review of our weak-continuous-measurement QT protocol for the general reconstruction of a density matrix. In Sec. III, we specialize to the case of atomic spin systems, and in particular, the control of the ground-electronic manifold of cesium atoms. Two cases are studied for different control methods: (1) QT on the 7-dimensional $F=3$ manifold of Cs controlled by the quadratic AC-Stark shift and quasi-static magnetic fields; (2) QT on the full 16-dimensional $F=3\oplus F=4$ electronic ground state subspace, controlled by time dependent radio frequency and microwave magnetic fields. In the former case, experiments have been carried out and we review those results.  The latter case is much more ambitious and we outline the challenges as shown from our simulations.  Finally, Sec. IV includes the conclusions and outlook of this work.

\section{Quantum tomography via continuous measurement protocol}\label{sec:QTprotocol}

\subsection{The basic protocol}\label{sec:QTprotocol:basic}
The general setting for our protocol is as follows.  One is given an ensemble of $N$, noninteracting, simultaneously prepared systems in an identical state $\rho_0$ that can be controlled and probed collectively.  We seek to find an estimate of the state of the system by continuously measuring some observable $\cO_0$. We restrict our attention to states in Hilbert spaces of finite dimension $d$ and measure traceless-Hermitian observables in the algebra $\mathfrak{su}(d)$.  In an idealized form, the probe performs a QND measurement that couples uniformly to the collective variable across the ensemble and measures $\sum_j^N \cO_0^{(j)}$.  For a sufficiently strong QND measurement, quantum back-action will result in substantial entanglement between the particles.  For example, such a phenomenon has been employed to create spin squeezed states of an ensemble when the fluctuations in projection-valued measurements (``projection noise") can be seen resolved within fundamental quantum fluctuations in the probe (``shot noise") \cite{kuzmich2000, appel2009, takahashi2009, vuletic2010, mitchell2010}.  We consider the opposite case of a very weak measurement such that back-action noise is negligible compared with the detector noise.   In this case, the procedure can be analyzed as a single atom control problem in which each member of the ensemble evolves under the same dynamics with state $\rho(t)$.  In this case, the measurement record is proportional to
\begin{equation}
M(t)=\Tr(\cO_0 \rho(t))+\sigma W(t),
\label{eq:measurementconti}
\end{equation}
amplified by the total number of atoms.  Here, $W(t)$ is Gaussian-random variable with zero mean and variance $\sigma^2$, that is introduced to account for the noise on the detector. 

In order to generate a measurement record that can be inverted to determine the initial state, one must control the dynamics so as to continuously write new information onto the measured observable.  To do so, the system is manipulated by external fields.  The Hamiltonian of the system, $H(t)=H[\phi_i(t)]$, is a functional of a set of time-dependent control functions, $\phi_i(t)$, which are chosen so that the dynamics produces an informationally complete measurement record $M(t)$.  Since our objective is to estimate the initial state of the system from the measurement record and  our knowledge of the system dynamics, it is more convenient to carry out the procedure in the Heisenberg picture.  Expressed this way, the state is fixed and control is used to generate new observables that we measure.  Note, this is generally different from the standard Heisenberg picture in that we allow for decoherence during the dynamical evolution.  We will return to this issue below.   The measurement record, Eq. (\ref{eq:measurementconti}), is then written $M(t)=\Tr(\cO(t)\rho_0)+\sigma W(t)$. For implementation of the algorithm, a time discretization of the problem is necessary. We sample the measurement record at discrete times so that 
\be
M_i=\Tr(\cO_i\rho_0)+\sigma W_i.
\label{eq:measurement}
\ee 
We have thus reduced the problem of QT to a linear stochastic estimation problem.  The goal is to determine $\rho_0$ given $\{M_i\}$ for a well chosen $\{\cO_i\}$ in the presence of noise $\{W_i\}$.

A number of transformations are necessary to increase the numerical stability and reliability of the algorithm. Let $\{E_{\alpha}, I/\sqrt{d}\}$, $\alpha=1,\dots, d^2-1$, be an orthonomal Hermitian basis of matrices, where $I$ is the identity matrix and $\Tr(E_{\alpha})=0$. The unknown initial state, $\rho_0$, can thus be decomposed as
\be
\rho_0=\frac{1}{d}I+\sum_{\alpha=1}^{d^2-1}r_{\alpha}E_{\alpha},
\label{eq:parameterizedrho}
\ee 
where $r_{\alpha}= \Tr(\rho_0 E_{\alpha})$ are real numbers.  We can then write Eq. (\ref{eq:measurement}) as
\be
\begin{split}
M_i&=\sum_{\alpha=1}^{d^2-1}r_{\alpha}\Tr(\cO_iE_{\alpha})+\sigma W_i,~~\text{or,}\\
\vec{M}&=\mbf{\tilde{\cO}}\mbf{r}+\sigma\vec{W},
\end{split}
\label{eq:measurement2}
\ee 
which in general is an overdetermined set of linear equations with $d^2-1$ unknowns $\mbf{r}=(r_1,\dots,r_{d^2-1})$ and where $\tilde{\cO}_{i\alpha}=\Tr(\cO_iE_{\alpha})$.

Eq. (\ref{eq:measurement2}) explicitly states that the conditional probability of the random variable $\vec{M}$ given the state $\vec{r}$ is the Gaussian distribution
\be
\mathcal{P}(\vec{M}|\vec{r})\propto \exp{\left(-\frac{1}{2\sigma^2}(\vec{M}-\mbf{\tilde{\cO}}\vec{r})^T(\vec{M}-\mbf{\tilde{\cO}}\vec{r})\right)}.
\label{eq:measurementdistribution}
\ee
We can use the fact that the argument of the exponent in Eq. (\ref{eq:measurementdistribution}) is a quadratic function of $\vec{r}$ to write the likelihood function
\be
\mathcal{P}(\vec{M}|\vec{r})\propto \exp{\left(-\frac{1}{2}(\vec{r}-\vec{r}_{ML})^T\vec{C}^{-1}(\vec{r}-\vec{r}_{ML})\right)},
\label{eq:likelihoodfunction}
\ee
describing a Gaussian function over possible states $\vec{r}$ centered around the most likely state, $\vec{r}_{ML}$, with covariance matrix, $\mbf{C} = \sigma^2(\mbf{\tilde{\cO}}^T\mbf{\tilde{\cO}})^{-1}$.  The unconstrained maximum likelihood solution is given by 
\be
\mbf{r}_{ML}=(\mbf{\tilde{\cO}}^T\mbf{\tilde{\cO}})^{-1}\mbf{\tilde{\cO}}^T\mbf{M}.
\label{eq:LSsolution}
\ee 
Since we treat the noise as Gaussian, this solution corresponds to the least squares solution of the linear system in Eq. (\ref{eq:measurement2}),  \cite{tarantola05}.   

The eigenvectors of $\mbf{C}^{-1}$ specify the directions in the operator space $\mathfrak{su}(d)$ that have been measured and its eigenvalues are the square of the signal-to-noise ratio of those measurements.  The covariance matrix thus allows us to quantify the information extracted from the measurement record.  A sufficient condition for an informationally complete measurement record (though not necessary unit fidelity due to noise) is one for which $\mbf{C}^{-1}$ is full rank.  This will be true when $\{\cO_i\}$ spans $\mathfrak{su}(d)$.  To achieve an informationally complete measurement record, the quantum system must be {\em controllable} in the sense that we can map any $\cO_0$ to any $\cO_i$ over the Lie algebra.  At early times during the reconstruction procedure, $\mbf{C}^{-1}$ will not be full rank and thus we use the Moore-Penrose pseudo inverse in Eq. (\ref{eq:LSsolution})  \cite{ben-israel03}.  Also, it is essential that the dynamics be sufficiently coherent such that an informationally complete set of observables can be generated before decoherence erases the state.  

\subsection{The positivity constraint}
For an informationally incomplete measurement record and/or for finite noise, the unconstrained maximum likelihood solution, Eq. (\ref{eq:LSsolution}), generally produces estimates of the density matrix with negative eigenvalues. To obtain a physical estimate of the state, we must impose the constraint that the estimated density matrix be positive semidefinite.  Such a constraint can be enforced through an appropriate parametrization of the unknown initial state, e.g., $\rho_0=T^\dg T$ where $T$ is a lower diagonal matrix \cite{paris04, klose01}. Although this parametrization has the advantage that the estimated state is Hermitian and positive-semidefinite by definition, it is not compatible with our continuous measurement protocol.  A least squares solution to Eq. (\ref{eq:measurement}) would involve a nonlinear unconstrained optimization for which there is no known efficient solution.  

We thus turn to constrained numerical optimization to find the ``closest" positive matrix to the unconstrained-maximum-likelihood estimate, i.e., the constrained-maximum-likelihood estimate, $\bar{\rho}$.  The covariance matrix determines a natural cost function metric with which to measure the distance these states according to
\be \label{E:costnorm}
	\| \mathbf{r}_{\rm ML}-\bar{\mathbf{r}}\|^2 = (\mathbf{r}_{\rm ML}-\bar{\mathbf{r}})^T \mathbf{C}^{-1}(\mathbf{r}_{\rm ML}-\bar{\mathbf{r}}) .
\ee
Technically speaking, this quantity is not a norm but rather a seminorm only when informationally incomplete measurements are considered ($\vec{C}$ is not full-rank), meaning that there exist some vectors $\mathbf{v}$ such that $\| \mathbf{v} \| = 0$ but $\mathbf{v} \not=0$ in those cases.  The use of this metric can be justified as follows. The inverse of the covariance matrix, $\mbf{C}^{-1}$, encodes all of the information about the independent directions in operator space that are being measured by our procedure.  A small eigenvalue of  $\mbf{C}^{-1}$ means a low signal-to-noise ratio associated with measurements of the corresponding eigen-operator, and thus that little is known about the trace-projection of the initial state onto that operator direction. The cost function, Eq. (\ref{E:costnorm}), takes into account that different directions in the space $\mathfrak{su}(d)$ are not measured in the same way and weights this in the distance between the initial estimate and the positive state.  In this way, during the numerical optimization, the more uncertain components of $\bar{\rho}$ can be adjusted more freely than the more certain ones, thereby maintaining faithfulness with the measurement record, but ensuring positivity.   To find the physical estimate we thus solve the following optimization problem:
\be
\rm{minimize}~~ \| \mathbf{r}_{\rm ML}-\bar{\mathbf{r}}\|^2
\label{eq:convexprogram}
\ee 
subject to
\be
\frac{1}{d}I+\sum_{\alpha=1}^{d^2-1}\bar{r}_{\alpha} E_{\alpha}\ge 0.
\label{eq:convexprogramconstraint}
\ee
While there is generally no analytic solution to this problem, it takes the form of a standard convex program since the matrix $\mbf{C}^{-1}$ is positive semidefinite and both the objective and the constraint are convex functions \cite{boyd08}.  The optimization is a semidefinite program which is efficiently solvable numerically. 
We implement this in MATLAB using freely available convex optimization package, CVX \cite{cvx10}.

\subsection{Control and dynamics}
An important question that has not yet been addressed is the way in which we drive the dynamics to generate the measurement record.  As discussed above, a sufficient condition is that the dynamics generate an informationally complete set of observables $\{\cO_i\}$, meaning that they span the Lie algebra $\mathfrak{su}(d)$.  The quantum dynamics must thus be ``controllable" in the Lie algebraic sense \cite{schirmer02}.  The Hamiltonian that governs the dynamics is a functional of a set of control waveforms such as externally applied fields parametrized by frequencies, amplitudes, and phases.  Our task is then to chose these waveforms to generate an  informationally complete set of observables $\{\cO_i\}$ in the desired time.   In practice, we fix the duration of the measurement record as determined by the characteristic time scales for evolution, dictated by both the Hamiltonian evolution for the given power in the controls and decoherence.  We choose the total time $T$ to be such that we can attain a good approximation to any unitary evolution matrix in $SU(d)$.  The total time is then coarse-grained into slices of duration $\delta t$, consistent with the slew rates and bandwidth constraints of the waveform drivers in the laboratory.  We thus reduce the problem to specification of a discrete set of waveform values compatible with experimental constraints.  The translation of the discretely sampled parameters to the continuous-time waveform depends on the characteristics of the physical drivers and the challenges of numerical integration, to be discussed below.

With the Hamiltonian in hand, we must choose the control parameters.  There is no unique solution; any choice that yields an informationally complete set $\{\cO_i\}$ in the given time series will suffice.  In principle, one would like to optimize the information gain over time $T$, and yield an unbiased state reconstruction.  This amounts to optimizing the entropy associated with the eigenvalues of the covariance matrix.  We have found empirically that the landscape for performing such an optimization is not favorable, and this approach becomes intractable, even for moderately sized Hilbert spaces ($d>9$).  Instead, our numerical studies show that one can achieve the required high fidelity and unbiased measurement record by choosing the control parameters {\em randomly} and unbiased over a designated interval.  We will demonstrate this below for the specific example of control and measurement of atomic hyperfine spins.  A more rigorous justification of this approach is still under consideration.  We have previously seen a connection between evolution via random unitary dynamics and the generation of an informationally complete measurement record \cite{merkel10}.   This may give us clues to optimally designing the control waveforms.

An essential component of this protocol is accurate modeling of the dynamical evolution of the observables measured in the continuous signal.  Fundamental to this is decoherence induced by the environoment.  Under typical conditions of Markovian evolution, these dynamics are generated by a Lindblad master equation,
\begin{eqnarray}
\frac{d\rho(t)}{dt} & =& \mathcal{L}_t [\rho(t)]  \nonumber \\
&=& -i [H(t), \rho(t)] -\frac{1}{2}\sum_\mu \left( L_\mu^\dg L_\mu \rho(t) + \rho(t) L_\mu^\dg L_\mu  \right) \nonumber \\
&+& \sum_\mu L_\mu  \rho(t) L_\mu^\dg.
\label{eq:mastereq}
\end{eqnarray}
The formal solution to this equation is a completely positive map on the initial density operator, $\rho(t) = \mathcal{V}_t [\rho(0)]$ with $\mathcal{V}_t$ being the solution to 
\be
\frac{d \mathcal{V}_t }{dt} = \mathcal{L}_t   \mathcal{V}_t \Rightarrow  \mathcal{V}_t = \mathcal{T} \left(\exp \int_0^t \mathcal{L}_s ds \right),
\ee
where $\mathcal{T}$ is the time-ordering operator.

We seek, however, the solution to the {\em Heisenberg} evolution, given formally by the adjoint map $\cO(t)=\mathcal{V}_t^\dag[\cO(0)]$, satisfying 
\be
\frac{d \mathcal{V}_t^\dag}{dt} =   \mathcal{V}_t^\dag \mathcal{L}_t ^\dag.
\label{adjoint}
\ee
Naively, one might assume that the generalization of the master equation for $\rho(t)$, Eq. (\ref{eq:mastereq}), to the Heisenberg evolution for $\cO(t)$ is
\begin{eqnarray}
\frac{d\cO(t)}{dt}   & =& \mathcal{L}_t^\dg [\cO(t)]  \nonumber \\
&=& +i [H(t),\cO(t)] -\frac{1}{2}\sum_\mu \left(  \cO(t) L_\mu^\dg L_\mu + L_\mu^\dg L_\mu \cO(t) \right) \nonumber \\
&+& \sum_\mu L_\mu^\dg \cO(t) L_\mu.
\end{eqnarray}
However, this is {\em not generally true} since $\mathcal{V}_t^\dag \circ \mathcal{L}_t \ne  \mathcal{L}_t \circ \mathcal{V}_t^\dag$.  Note that the correct Heisenberg evolution is
\be
\frac{d \cO(t)}{dt} =   \mathcal{V}_t^\dag \left[ \mathcal{L}_t ^\dag[\cO(0)] \right],
\ee
and $d \cO(t)/dt \ne  \mathcal{L}_t ^\dag[\cO(t)]$ unless $\mathcal{L}$ is time independent.  The lack of commutativity between the adjoint map and its generator will be the case for a generic time-dependent control Hamiltonians under consideration here. Because of this, the decohering Heisenberg operators do not satisfy a {\em time-local differential equation} \cite{breuer03}.  This severely complicates the efficiency with which we can integrate the dynamics to determine the measurement set $\{\cO_i\}$.  

To deal with this problem in moderately large Hilbert spaces, as we will discuss in Sec. \ref{sec:CsRFUW}, we restrict our waveform so that the control parameters are {\em piecewise constant} over a reasonable duration.  Then, over each interval we can simply exponentiate the Lindbald generator of the superoperator map.  The traceless operators, $\mathcal{O}_i$, are ``vectorized" to a large column of dimension $d^2-1$ and the superoperator, $\mathcal{V}_{t_i}$, is a large $(d^2-1)\times(d^2-1)$ matrix expanded in the basis $E_{\alpha}$.  Using curved bra-(row) ket-(column) notation for the supervectors and superoperators, our integration then takes the form
\be
\left(\cO_i \right| = \left(\cO_0 \right| \mathcal{V}_{t_i},
\label{eq:supopevolutionO}
\ee
where 
\be
\mathcal{V}_{t_{i+1}} = e^{\mathcal{L}_{t_i}\delta t}\, \mathcal{V}_{t_i}.
\label{eq:supopevolutionV}
\ee
For non-piecewise constant controls, this corresponds to an Euler integration of the completely positive map.  Such an approximation will be very inefficient and numerically unstable for large dimensional systems.  For this reason a piecewise constant control is best suited to our protocol.

\subsection{Further technical considerations}

Beyond decoherence, an essential ingredient for accurate modeling of the dynamics is parameter estimation. The ability to reach high fidelities for the estimated states relies on the assumption that we know exactly how the system is evolving at the time that the data is taken so that we know exactly which operators $\{\cO_i\}$ are being measured.  This means that all of the parameters in the Hamiltonian must be precisely known before the quantum state reconstruction is even possible.  In practice, many such parameters can be precalibrated.  However, other parameters, such as a background fields may be unknown, and other parameters may be inhomogeneous across the ensemble.  Our protocol is robust because, unlike other optimal control tasks, such as state-to-state mapping, the exact parameters of the experiment need not be fixed.  As long as we can determine, post-priori, the operating conditions via parameter estimation, and the signal is informationally complete, we can extract the quantum state with high fidelity.  

An additional parameter we must fix is the initial observable being measured.  Though abstractly we have called this $\cO_0$, in practice the true observable may delicately depend on special alignment of the apparatus.  We will see how we can use parameter estimation as well to fix this observable and the overall calibration of the signal in physical units when compared with the dimensionless units treated here.  

Finally, one technical detail that we have not discussed so far is the signal-to-noise ratio (SNR).  Even under ideal conditions, the fidelity of the QT is fundamentally limited by noise.  For Gaussian white noise, it is essential to limit the bandwidth in which we analyze the measurement record and the dynamics must be chosen so that the relevant information about the state is contained in a narrow frequency band. In addition, $1/f$ noise in the detector dictates that the relevant signal be sufficiently far from DC.  To maximize our SNR for a given ensemble, we pass the measurement signal through a narrow bandpass filter.  In any numerical simulation of the expected fidelity, the simulated signal is passed through the equivalent digital filter as used in the laboratory.  This ensures equivalent measurement records for equivalent operating conditions.  

In the next section, we will apply our QT protocol to the reconstruction of states encoded in atomic hyperfine spins.  After defining the system, we can simulate a measurement record in the presence of noise, decoherence and errors for a given initial state $\rho_0$ and use it to run it through the algorithm to find the estimate $\bar{\rho}$.  In order to quantify the performance of our method, we calculate the fidelity
\be
\mathcal{F}(\bar{\rho},\rho_0)=\left[\rm{Tr}\left(\sqrt{\sqrt{\bar{\rho}}\rho_0\sqrt{\bar{\rho}}}\right)\right]^2.
\label{eq:fidelity}
\ee
Good performance is judged by high fidelity averaged across a collection of randomly sampled states.

\section{Application to hyperfine spins}\label{sec:Cs}
\begin{figure}[t]
\begin{center}
\includegraphics[width=8.5cm,clip]{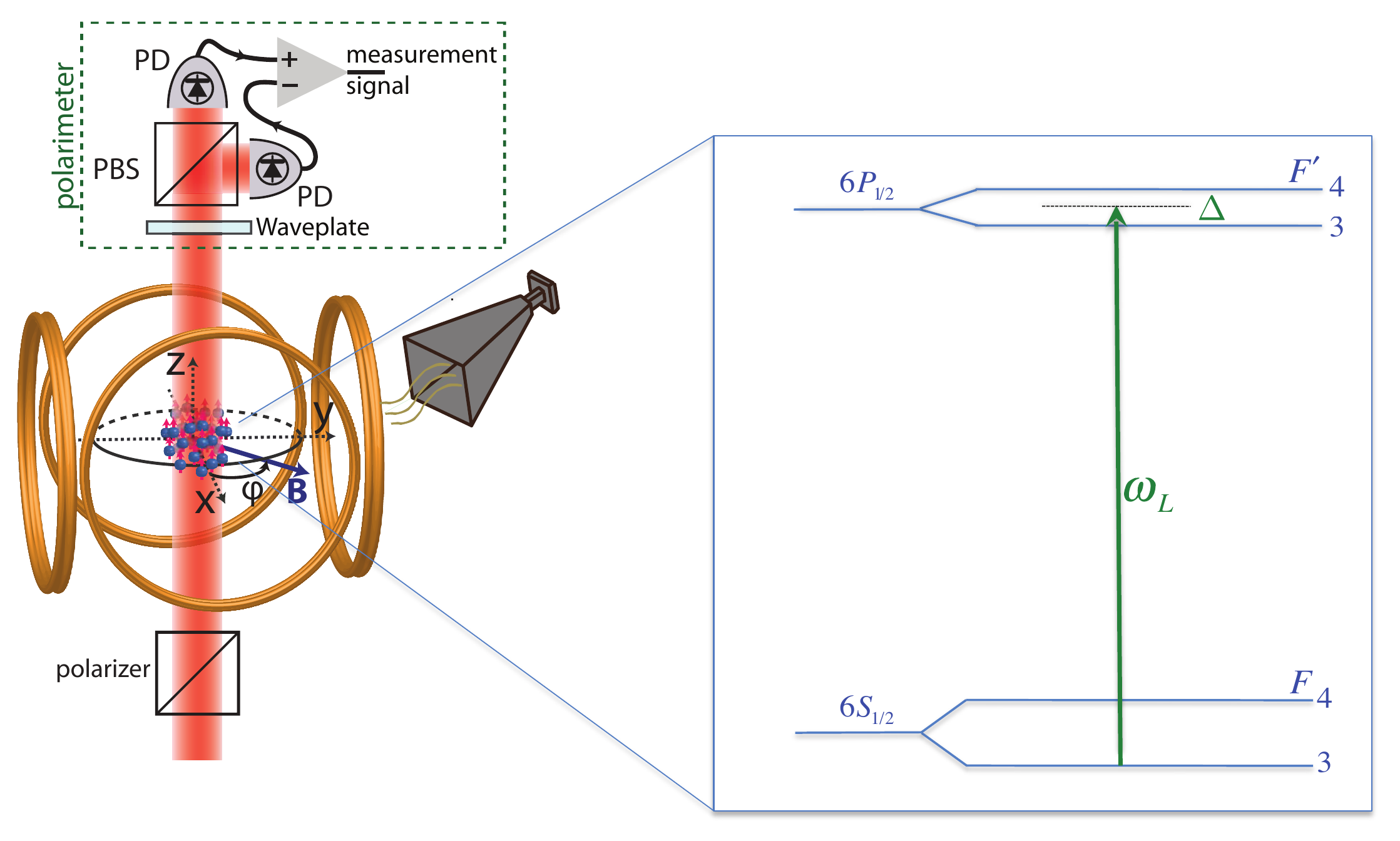}
\caption{(Color online) Schematic of our system geometry.  A cold gas of atoms is collected from a magneto-optic-trap/optical molasses, and optically pumped to form a nearly pure ensemble of identical spins.  The spins are controlled through a combination of light-shift interaction, magnetic fields produced by pairs of Helmholtz coils, and microwave fields.  A measurement of the spins is performed by polarization analysis of the transmitted probe.  A sketch of the atomic level structure for $^{133}$Cs is shown inset (not to scale).}
\label{F:setup}
\end{center}
\end{figure}

In this section, we apply the QT protocol to the reconstruction of states encoded in spin systems. Our platform is the hyperfine manifold of magnetic sublevels associated with the ground-electronic state of laser-cooled alkali-metal atoms, providing a Hilbert space of dimension $d=(2S+1)(2I+1)$ where $S=1/2$ is the single valence electron spin and $I$ is the nuclear spin.  In particular, we work with $^{133}$Cs, whose nuclear spin is $I=7/2$, yielding hyperfine coupled spins of magnitude $F=3,4$ and a total Hilbert space of dimension $d=16$.  A schematic of the system, including atomic level structure, control, and measurement components, is shown in Fig. (\ref{F:setup}).

The spins can be controlled through a combination of magnetic interactions and off-resonant optical coupling from a laser field.  The fundamental Hamiltonian is
\be
H(t) = A\,\mbf{I}\cdot\mbf{S} -\boldsymbol{\mu}\cdot\mbf{B}(t) -\frac{1}{4}E_i^*  \alpha_{ij} E_j,
\label{eq:controlH}
\ee
where $A$ is the hyperfine coupling constant,  $\boldsymbol{\mu}$ is the atomic magnetic moment operator, and $\alpha_{ij}$ is the atomic dynamic polarizability operator, both depending on the atomic spin degrees of freedom.   Here and throughout, we take the laser field complex amplitude, $\mbf{E}$, to be fixed, and control is accomplished through time-variation of the magnetic field, $\mbf{B}(t)$.  Under typical operating conditions where the hyperfine coupling dominates over all other forces, the total spin angular momentum, $F=I\pm1/2 =3,4$ and its projection along a quantization axis, $m$, are approximate good quantum numbers and define the basis of states in the Hilbert space we seek to control.  

A central component of our QT protocol is quantum controllability.  A finite dimensional system with a generic Hamiltonian of the form $H(t) =  \sum_j^k \lambda_j(t) H_j$, with external fields determined by $\lambda_j(t)$, is said to be controllable if $\{H_j\}$ generate the Lie algebra of the relevant group of unitary matrices on the space \cite{schirmer02}.  As we generally do not measure the trace of the density operator, we restrict our attention to the Lie algebra $\mathfrak{su}(d)$.  For situations in which we seek control in a single irreducible manifold $F$, the relevant algebra is $\mathfrak{su}(2F+1)$, $F=3,4$; on the the entire hyperfine manifold the algebra is $\mathfrak{su}(16)$.  For $F>1/2$, this requires Hamiltonians that are not linear in all of the components of $\mbf{F}$.  A challenge for any implementation is to  access nonlinear interactions that render the system controllable and coherent.  

Continuous measurement of the system is carried out through polarization spectroscopy of a probe laser beam that passes through the ensemble while it is being controlled.  The atoms induce a polarization dependent index of refraction in a manner depending on their spins' state according to the light-shift interaction \cite{deutsch09}.  In the limit of negligible backaction, the effect of the interaction is a rotation of the probe's Stokes vector $\mbf{S}$ on the Poincar\'{e} sphere according to the rotation operator $U_R = \exp \left( -i \chi_0 \avg{\mbf{\mathcal{O}}} \cdot \mbf{S} \right)$, where $\chi_0= OD_0 (\Gamma/2\Delta_c)$ is the characteristic rotation angle depending on the resonant optical density, $OD_0$, and a characteristic detuning from resonance, $\Delta_c$.  Taking the $z$-axis along the direction of propagation of the probe, the components of the vector of atomic observables that generate the rotations about the three axes of the Poincar\'{e} sphere are,
\begin{subequations}
\begin{align}
\vec{\mathcal{O}}\cdot\vec{e}_1 &= \sum_{F,F'} C^{(2)}_{F'F} \frac{\Delta_c}{\Delta_{F'F}}\left( \frac{F_x^2 - F_y^2}{2} \right) \\
\vec{\mathcal{O}}\cdot\vec{e}_2 &= \sum_{F,F'} C^{(2)}_{F'F} \frac{\Delta_c}{\Delta_{F'F}}\left( \frac{F_x F_y + F_y F_x}{2} \right) \\
\vec{\mathcal{O}}\cdot\vec{e}_3 &= \sum_{F,F'} C^{(1)}_{F'F} \frac{\Delta_c}{\Delta_{F'F}} F_z 
\label{eq:FaradayObservable}
\end{align}
\end{subequations}
where $C^{(K)}_{F'F}$ are coupling constants that depend on the irreducible rank-$K$ tensor polarizability for the given probe detuning $\Delta_{F'F}$ from the ground $(nS_{1/2})F$  to the excited $(nP_J')F'$ manifold \cite{deutsch09}.  For weak interactions under consideration here, $\chi_0 \ll 1$, this rotation corresponds to a small local displacement.  Measurement of the Stokes vector component along the direction $\hat{n}$ then correlates with a measurement of the atomic operator $\hat{n}\cdot \mbf{\mathcal{O}}$.  Preparing the probe initially linearly polarized along the $\mbf{e}_1$ of the Poincar\'{e} sphere, and analyzing along the direction, $\mbf{n} = \cos{\theta}\mbf{e}_2+\sin{\theta} \mbf{e}_3$, the general measurement record will be of the form
\be
M(t)= a\avg{F_x F_y + F_y F_x}_t+b \avg{F_z}_t+ \sigma W(t),
\label{eq:fullrecord}
\ee
where $a$ and $b$ are constants that depend on the vector and tensor contributions to the polarizability for the given detuning, as well as the polarization the analysis direction, $\theta$.  

The light-shift interaction also induces dynamics on the atomic spin, depending on the probe's polarization. The combination of coherent evolution and decoherence due to photon scattering can be modeled by a master equation of the form \cite{deutsch09}
\be
\begin{split}
\frac{d\rho(t)}{dt}&=-i\left(H_{\rm eff}(t)\rho(t)-\rho(t)H_{\rm eff}^{\dg}(t)\right)\\
                             &+\Gamma\sum_{q}\left(\sum_{F,F_1}W_q^{F F_1}\rho^{F_1 F_1}(t) W_q^{F F_1\dg}\right.\\
                             &+\left.\sum_{F_1\ne F_2}W_q^{F_2 F_2}\rho^{F_2 F_1}(t) W_q^{F_1F_1\dg}\right).
\end{split}
\label{eq:csmaster}
\ee
In this equation, projections of operators onto subspaces with a given $F$ are denoted, $A^{F_1 F_2} = P_{F_1} A P_{F_2} = \sum_{m_{1} m_{2}} \ket{F_1,m_{1}}\bra{F_1,m_{1}}A \ket{F_2,m_{2}}\bra{F_2,m_{2}}$.  The total effective Hamiltonian is given by $H_{\rm eff}(t) = H_{HF}+H_B(t)+H^{LS}_{\rm eff}$, where $H_{HF}$ is the hyperfine interaction, $H_B$ is the magnetic interaction, and the effective (non-Hermitian) Hamiltonian accounting for light-shift and optical pumping is
\be
H^{LS}_{\rm eff}=\frac{\Omega^2}{4}\sum_{FF'}\frac{(\boldsymbol{\epsilon}^*\cdot\vec{D}_{FF'})(\vec{D}_{F'F}^{\dg}\cdot\boldsymbol{\epsilon})}{\Delta_{F'F}+i\Gamma/2}.
\label{eq:lshamiltonian}
\ee
Here, $\Omega$ is the laser Rabi frequency for a unit oscillator strength.  The strength of the transitions for  $\boldsymbol{\epsilon}$-polarized light are accounted for by the dimensionless dipole raising operator,
\be
\vec{e}_q \cdot \vec{D}_{F'F}^{\dg}=\sum_m \mathcal{K}_{JF}^{J'F'}\braket{F'm+q}{Fm;1q}\ketbra{F'm+q}{Fm}
\ee
where the coefficient $\mathcal{K}_{JF}^{J'F'}$ is given in terms of Wigner 6j symbol
\be
\mathcal{K}_{JF}^{J'F'}=(-1)^{F'+I+J'+1}\sqrt{(2J'+1)(2F+1)}\left\{\begin{array}{ccc}F' & I & J'\\ J & 1 & F\end{array}\right\}.
\ee
The Lindblad jump operators are given by
\be
W_q^{F_b F_a}=\sum_{F'}\frac{\Omega/2}{\Delta_{F'F_a}+i\Gamma/2}(\vec{e}_q^*\cdot\vec{D}_{F_b F'})(\vec{D}_{F'F_a}^{\dg}\cdot\boldsymbol{\epsilon}),
\label{eq:jump}
\ee
describing absorption of a photon with polarization $\boldsymbol{\epsilon}$, emission of a photon with polarization $q$, and optical pumping between hyperfine manifolds $F_a$ and $F_b$.  Transfer of population between sublevels by optical pumping occurs at a rate $\gamma_{F_a m_a \rightarrow F_b m_b} = \sum_q |\bra{F_b m_b} W^{F_b F_a}_q \ket{F_a m_a}|^2$.  The final term in the master equation, Eq. (\ref{eq:csmaster}), proportional to $\rho^{F_2 F_1}$, represents transfer of coherences that may exist between hyperfine manifolds, but are preserved in spontaneous emission when the detuning of the light is sufficiently large.

With this general framework in hand, we have the tools necessary for our QT protocol: control and continuous measurement.  We  apply this in two examples, considering different control scenarios, to achieve quantum state reconstruction.  

\subsection{Reconstructing a single $F$-manifold via light-shift control}\label{sec:CsLightShift}

Previous work by \cite{silberfarb05} and \cite{smith06} demonstrated the QT protocol on the $F=3$ ground-state manifold of Cs.  In that case, the light acted simultaneously as the probe on the system and the driver of nonlinear spin dynamics through the light shift, and thus made the system  controllable on $SU(2F+1)$. For linear polarization of the laser probe along $x$, the effective light-shift Hamitonian Eq. (\ref{eq:lshamiltonian}) in a given manifold $F$ can be expressed in irreducible tensor components as \cite{deutsch09}
\begin{equation}
H_{{\rm eff},F}^{LS}=\gamma_{sc} \left[\left(\beta^{(0)}_F - \beta^{(2)}_F \frac{F(F+1)}{3} \right) I_F + \beta^{(2)}_F F_x^2 \right]
\label{eq:effhamiltonianls}
\end{equation}
where $I_F$ is the identity operator on the hyperfine manifold $F$ and
\begin{equation}
\beta^{(K)}_F =\frac{2 \Delta_c^2}{\Gamma^2} \sum_{F'}C^{(K)}_{F'F} \frac{\Gamma/2}{ \Delta_{F'F} +i \Gamma/2}
\label{eq:betas}
\end{equation}
are complex coupling coefficients depending on the rank-$K$ atomic polarizability.  The real part leads to the light shift and the imaginary part causes decoherence via photon scattering.  For emphasis, we have explicitly factored out the characteristic photon scattering rate $\gamma_{sc} = (\Omega^2  \Gamma)/(4 \Delta_c^2)$, which sets the time scale for dynamics on the atom-photon interaction.  The real part of $\beta^{(0)}$ is a shift independent of magnetic sublevel and thus can be neglected here.  The real part of $\beta^{(2)}$, however, is essential for controllability.  The best figure of merit for nonlinearity vs. decoherence is obtained if the laser is detuned approximately in between of the excited manifolds of the D1 line.  Precisely, we choose the detuning from $(6S_{1/2})F=3$ to $(6P_{1/2})F'=3$ to be $\Delta_c/2\pi=642.78$ MHz.  For this choice of detuning, the nonlinear light shift scales relative to $\gamma_{sc}$ as $\text{Re}(\beta^{(2)}) = 6.53$, while decoherence scales as $\text{Im}(\beta^{(0)}) = -0.23$ and $\text{Im}(\beta^{(2)}) = 0.005$, which implies sufficient coherent control can occur for QT before decoherence erases the information.

We can achieve full control of a hyperfine manifold $F$  through a combination of the nonlinear light shift and magnetic interaction, as given in the fundamental Hamiltonian, Eq. (\ref{eq:controlH}).  In the linear Zeeman regime, the magnetic moment restricted to this subspace is $\boldsymbol{\mu}_F = -g_F \mu_B \mbf{F}$, where $g_F$ is the Land\'e g-factor .  Fixing the  magnitude of $\mbf{B}$ and allowing its direction to vary in the $x$-$y$ plane, the control Hamiltonian is
\be
H_B(t) =\Omega_L(\cos{\phi(t)}F_x+\sin{\phi(t)}F_y),
\label{eq:controlhamiltonianls}
\ee
where $\Omega_L= g_F \mu_B B$ is the Larmor frequency.  The operators $\{F_x, F_y\}$ along with $ F_x^2$, from the light-shift interaction, form a minimal set of generators of the Lie algebra $\mathfrak{su}(2F+1)$ for any $F$.   

Generation of an informationally complete set of observables is achieved through the choice of the waveform $\phi(t)$, the angle of orientation.  The Heisenberg dynamics evolve the observables according to Eqs. (\ref{eq:supopevolutionO}) and (\ref{eq:supopevolutionV}), with the Lindbladian given in Eq. (\ref{eq:csmaster}).  As we restrict our attention to control of a state in the subspace $F=3$ and the probe is detuned close to the excited state hyperfine splitting in order to generate the nonlinear light shift, the population in $F=4$ is essentially invisible to the probe.  Any optical pumping is thus treated as a loss and the master equation, Eq. (\ref{eq:csmaster}), restricted to a single manifold, is not trace-preserving.  Our reconstruction algorithm is insensitive to this loss in that we fix the state to be normalized through the parametrization, Eq. (\ref{eq:parameterizedrho}). In \cite{silberfarb05}, the control waveform was designed through a local optimization of the covariance matrix entropy.  This was possible given the moderate size of the Hilbert space ($d=7$).  For large dimensional system, this approach is not tractable, and instead we empirically choose random waveforms, as we will discuss in Sec. \ref{sec:CsRFUW}.

\begin{figure}[t]
\begin{center}
\includegraphics[width=7cm,clip]{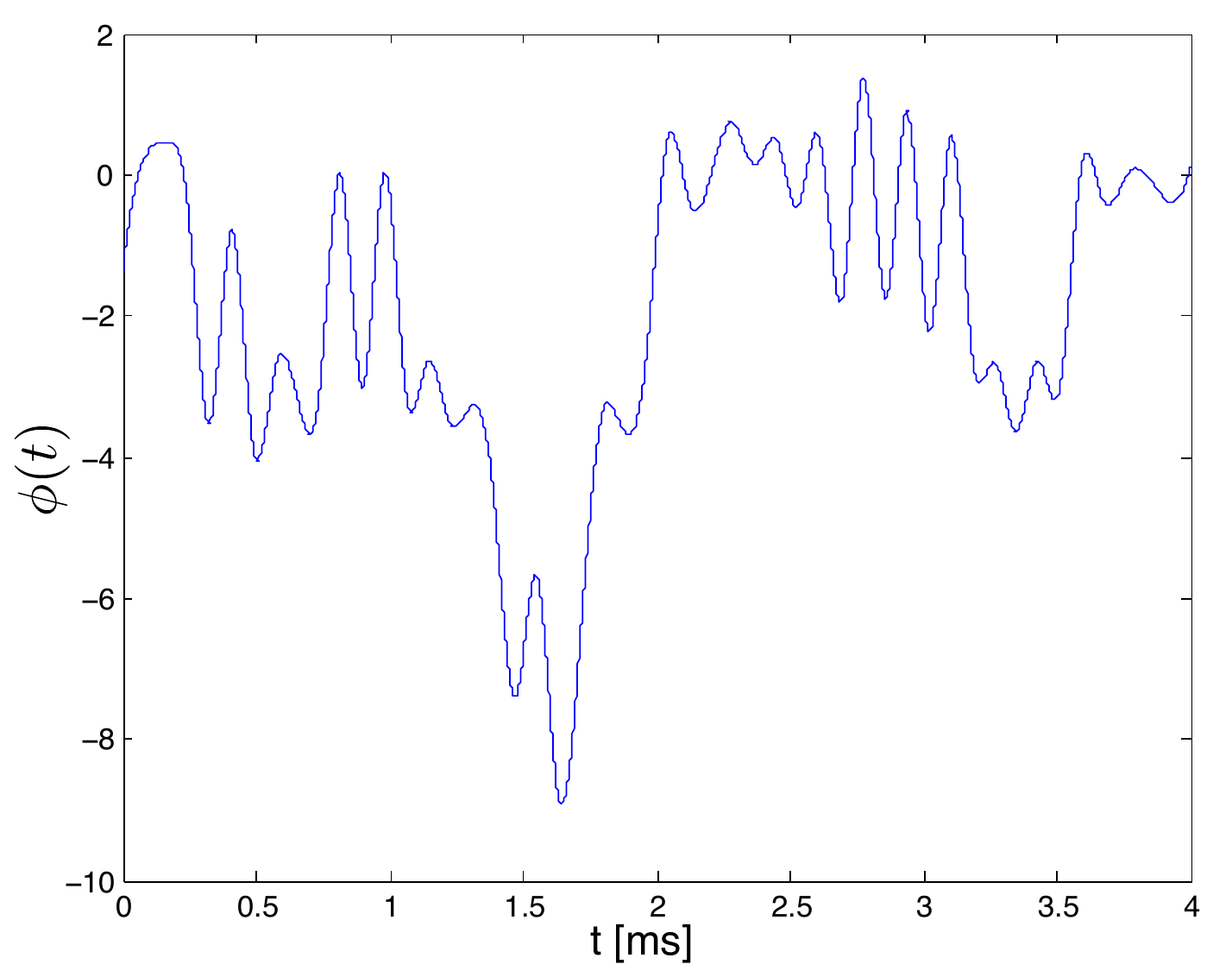}
\caption{(Color online) Control waveform as a function of time, $\phi(t)$, that determines the Zeeman Hamiltonian, Eq. (\ref{eq:controlhamiltonianls}). 50 discrete points of the function were optimized to maximize the amount of information acquaried while a cubic spline algorithm was used to interpolate them.}
\label{F:LScontrolangles}
\end{center}
\end{figure}

As an example of the performance of our quantum state reconstruction protocol, we show the reconstruction procedure for experimental data in \cite{smith06}.  The fields are chosen with nominal parameters $\Omega_L/2\pi$ = 17.5 kHz, $\gamma_{sc}/2\pi = 81.4$ Hz. For these time scales, full controllability is achieved in $\sim 4$ ms.  Over this duration, the waveform $\phi(t)$ is sampled by 50 control parameters $\phi_i$ at 80 $\mu$s intervals, consistent with the driver slew rates.  The optimized angle as a function of time, shown in Fig. \ref{F:LScontrolangles}, is made into a continuous waveform via cubic splines. With the time-dependent magnetic field interaction, and non-Hermitian light-shift Hamiltonian Eqs. (\ref{eq:controlhamiltonianls}) and (\ref{eq:effhamiltonianls}), respectively, the observables are evolved according to the adjoint of the map, Eq. (\ref{eq:supopevolutionV}).  In this case, the integration is an Euler approximation to the continuous-time differentiable Hamiltonian.  Again, because of the moderate size of the Hilbert space, this is possible.  For large dimensional systems, such integration is unstable and it is essential that the waveforms are piecewise constant.

\begin{figure}[t]
\begin{center}
\includegraphics[width=7cm,clip]{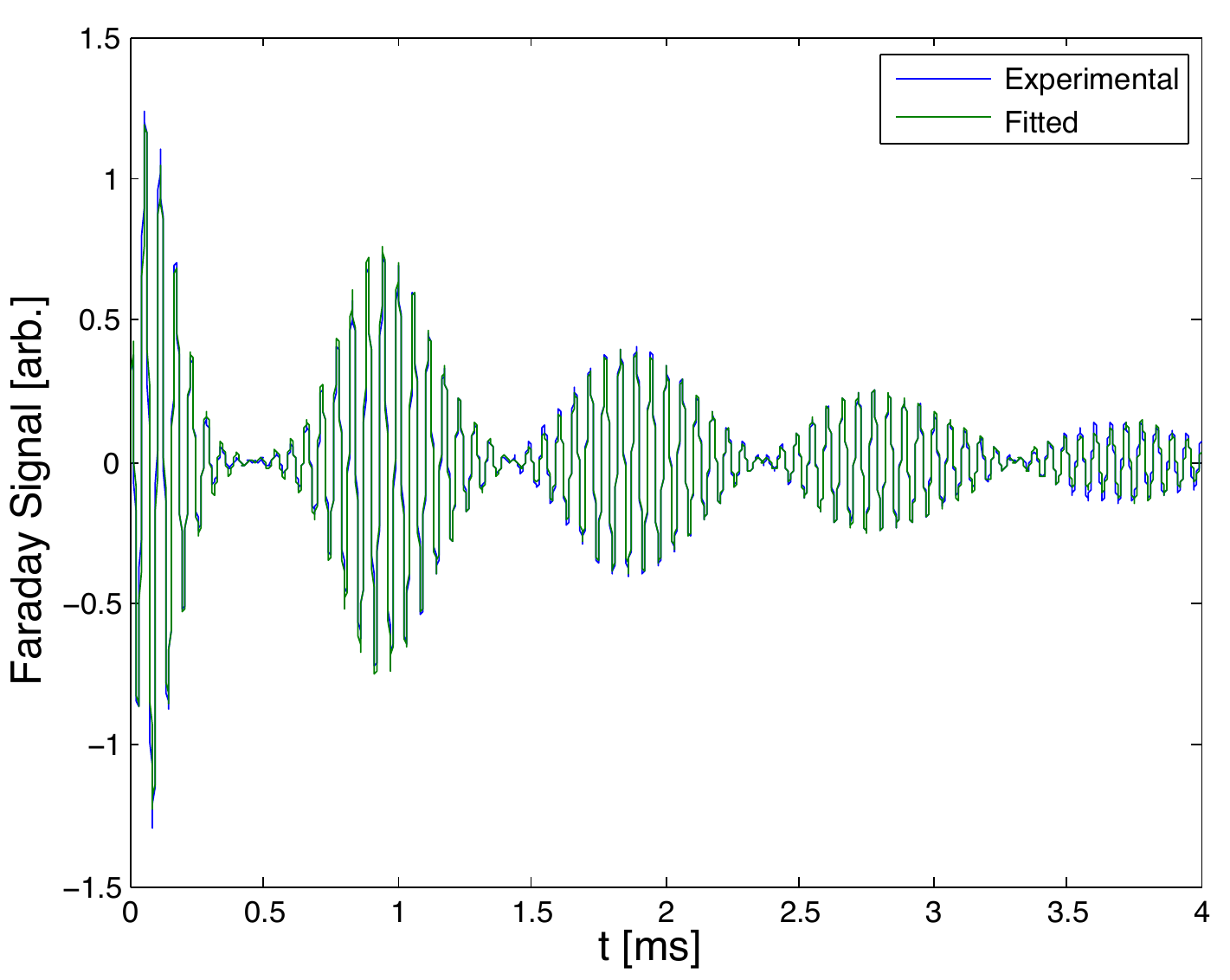}
\caption{(Color online) Comparison between the experimental Faraday rotation signal (blue) and the signal fitted by our model (green) for the case of Larmor precession in the presence of the nonlinear light-shift.  This signal is used to calibrate the intensity distribution seen by the atoms.}
\label{F:Larmor}
\end{center}
\end{figure}

Practical application of the protocol depends on accurate estimation of all of the parameters that define the experiment.  Before driving the atoms with the time-dependent control waveform that generates the informationally complete measurement record, we drive simple dynamics so that the Hamiltonian parameters can be calibrated. We prepare a spin coherent state, polarizing the atom along $y$ via optical pumping as our known initial state.  Fixing the control magnetic field along $x$, continuous measurement of atomic Larmor precession via Faraday rotation according the measurement record, Eq. (\ref{eq:fullrecord}), with $a=0$, is shown Fig. \ref{F:Larmor}.  The signal exhibits the collapse and revival of Larmor oscillations characteristic of the nonlinear spin dynamics first seen in \cite{smith04}.  Decoherence via photon scattering is exhibited by the decay of the overall signal.  In addition, the signal shows inhomogeneous broadening due to variations in the probe intensity across the ensemble.  Finally, technical issues such as finite response time affect the time origin of the measurement record.  All of these features must be accurately included in our model of the Heisenberg evolution in order to obtain high-fidelity QT. 

To perform parameter estimation, we employ a least-squares fit between the measured and simulated measurement record with cost function $\mathcal{C} = \| \mbf{M} - \mbf{\tilde{M}}(\Omega_L, \gamma_{sc}, t_0) \|^2$.  Here $\mbf{M}$ is the vector of time-sampled data from the calibration run and 
\be
\tilde{M}_i(\Omega_L, \gamma_{sc}, t_0) = \int \avg{\cO_i (\Omega_L, \xi \gamma_{sc}, t_0)} f(\xi) d\xi
\ee
is the simulated measurement time-series.  The unknown parameters are the Larmor frequency, $\Omega_L$, photon scattering rate, $\gamma_{sc}$, and origin of time, $t_0$.  In addition, we account for inhomogeneity in the laser intensity through the distribution function $f(\xi)$, where $\xi$  is the ratio between the nominal scattering rate and the local scattering rate at the position of the atom.  The parameter $\gamma_{sc}$ then represents the scattering rate at $\xi=1$ where $f(\xi)$ is peaked. Note, we take $f(\xi)$ to be unnormalized.  The overall scale determines the conversion between the simulated dimensionless signal and the laboratory measurement record.

In order to make this inversion tractable, we employ a linearly interpolated intensity distribution.   We parametrize  $f(\xi)$ as a piecewise linear function evaluated only at discrete points $\xi_n$,  $f(\xi)=m_n\xi+b_n$, for $\xi_{n-1}\le \xi<\xi_n$ where $m_n=\left[ f(\xi_n)-f(\xi_{n-1})\right]/\left(\xi_n-\xi_{n-1}\right)$ and $b_n=\left[\xi_n f(\xi_{n-1})-\xi_{n-1}f(\xi_{n})\right]/\left(\xi_n-\xi_{n-1}\right)$.  Making use of this parameterization, the simulated Larmor precession signal can be written 
\be
\begin{split}
\tilde{M}_i&=\sum_{n=1}^N\left\{\left(\frac{-Q_{n,i}+\xi_nT_{n,i}}{\xi_n-\xi_{n-1}}\right)f(\xi_{n-1})\right.\\
          &+\left.\left(\frac{Q_{n,i}-\xi_{n-1}T_{n,i}}{\xi_n-\xi_{n-1}}\right)f(\xi_n)\right\}
\end{split}
\ee
where $Q_{n,i}=\int_{\xi_{n-1}}^{\xi_{n}}\avg{\cO_i(\xi')}\xi d\xi'$ and $T_{n,i}=\int_{\xi_{n-1}}^{\xi_{n}}\avg{\cO_i(\xi')}d\xi'$. The measurement record can be expressed as a linear equation $\vec{M}=\vec{A}\vec{f}+\sigma\vec{W}$ where $\vec{f}=(f_0,f_1,\ldots,f_N)$ is the function $f(\xi)$ evaluated in $N+1=17$ discrete points. The least squares solution is $\vec{f}=(\vec{A}^T\vec{A})^{-1}\vec{A}^T\vec{M}$ and we use this result as our estimate of the intensity inhomogeneity. This procedure is iteratively repeated for different values of nominal intensity, magnetic field and origin of time until the cost function is minimized. 

\begin{figure}[t]
\begin{center}
\includegraphics[width=7cm,clip]{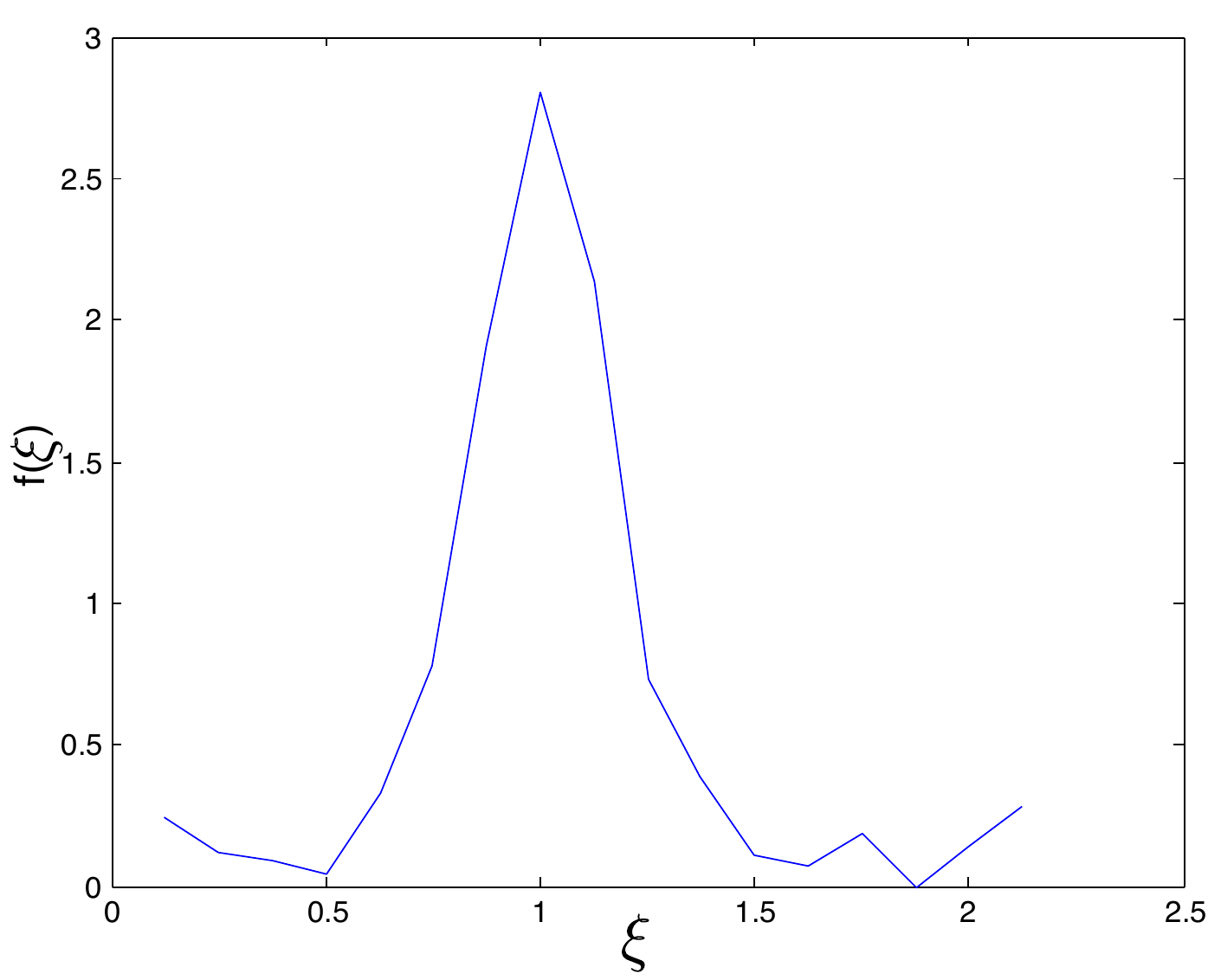}
\caption{(Color online) Estimated distribution of intensity over the atomic ensemble, $f(\xi)$, where $\xi$ is the ratio of the intensity seen by the atoms to the nominal (peak) intensity.}
\label{F:Distribution}
\end{center}
\end{figure}

Figure \ref{F:Larmor} shows a comparison between the experimental signal and the fitted one by the procedure described above. The estimated Larmor frequency and peak scattering rate for this plot are $\Omega_L/2\pi = 17.469$ kHz and $\gamma_{sc}/2\pi =84.7$ Hz. Figure \ref{F:Distribution} shows the piecewise linear estimation of the distribution of the intensity of the laser probe over the atomic ensemble. We use this distribution in the QT run of the experiment to average over the intensity inhomogeneity.

\begin{figure}[t]
\begin{center}
\includegraphics[width=7cm,clip]{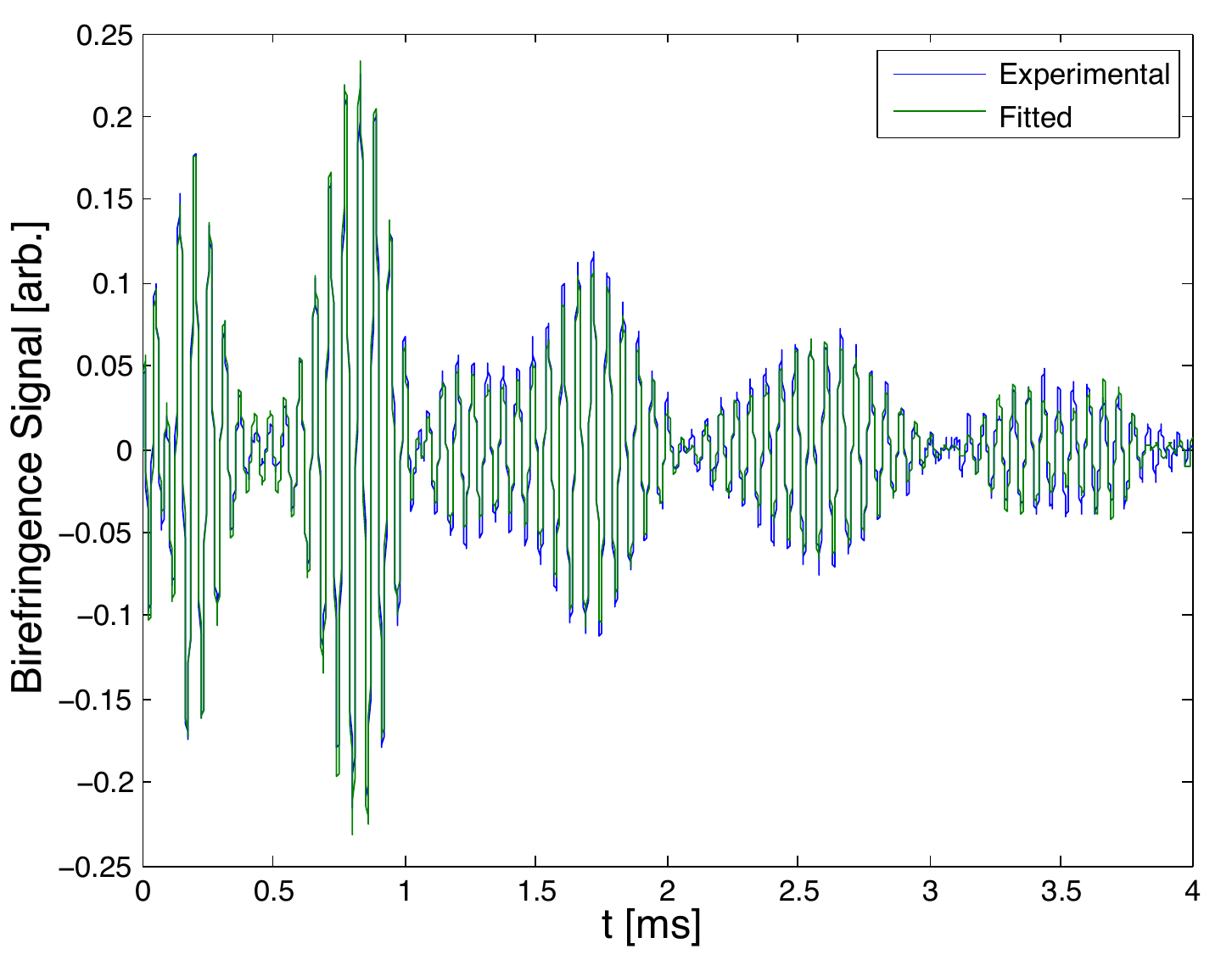}
\caption{(Color online) Comparison between the experimental birefringence signal (blue) and the signal fitted by our model (green) for the same Larmor precession dynamics given in Fig. \ref{F:Larmor}.  This signal is used to calibrate the measurement basis.}
\label{F:BirefringenceSignal}
\end{center}
\end{figure}

A final calibration must be performed to determine the exact observable we measure in the QT run.  For the protocol considered here, because we are necessarily detuned close enough to resonance in order to generate a sufficient nonlinear light shift, it is more advantageous to measure the birefringence of the atomic sample corresponding to the observable $F_x F_y + F_y F_x$.  This observable evolves more rapidly into the higher order polynomials in $\mbf{F}$ necessary to reconstruct the density operator as compared to $F_z$, measured in Faraday rotation.  Imperfections in waveplates, however, imply some uncertainty in the exact direction along the Poincar\'{e} sphere in which we perform a polarization analysis. We perform a second Larmor precession calibration run, but with a nominal quarter waveplate so that the polarization spectroscopy corresponds to a birefringence measurement.  According to Eq. (\ref{eq:fullrecord}), the actual measurement operator has a contamination of Faraday measurement.   Using the the estimated distribution of intensity obtained previously, $f(\xi)$, the simulated measured record in this case is $\tilde{M}_i=a\int\avg{F_xF_y+F_yF_x}_if(\xi)d\xi+b\int\avg{F_z}_i f(\xi)d\xi$. Note that in this expression, the unknown parameters are $a$ and $b$.  As before, these two parameters are obtained by a least squares fit in which we also fit the magnitude of the magnetic field and the origin of time.  For the data shown here $a=0.1613$ and $b=0.1598$.  Though the measurement corresponds almost completely to the $S_2$ direction on the Poincar\'{e} sphere, even a small Faraday contamination will produce a noticeable effect in the measurement record due to the fact that its measurement strength, being a rank-1 tensor effect, is higher than the birefringence's, which is a rank-2 tensor effect.  Accounting for the small misalignment is essential to guarantee high fidelity reconstruction.  The result of the simulated and measured signals for this calibration step are shown in Fig. \ref{F:BirefringenceSignal}.  The excellent agreement between our model and the measured signal allows us to achieve high fidelities in our QT procedure.

With calibrated parameters in hand, we carry out QT of a prepared quantum state. The system evolves under the Hamiltonian given in Eq. (\ref{eq:controlhamiltonianls}) for an appropriately chosen control wave form $\phi(t)$, as discussed above.  The effect of atomic birefringence on the transmitted probe is measured in our polarimeter and the resulting measurement record is modeled by Eq. (\ref{eq:fullrecord}) with the previously estimated parameters.  This record is then used in Eq. (\ref{eq:LSsolution}) to determine the maximum-likelihood estimate.  In addition, we carry out a final round of parameter estimation and fit for the magnetic field amplitudes in the two coils, $B_x$ and $B_y$, which produces slightly different Larmor frequencies in the $x$ and $y$ directions, and the origin of time, $t_0$, for the particular run. While the B-field is unlikely to change between the initial calibration and the QT run, we find that refitting is necessary to ensure high fidelity.  The origin of this discrepancy is unknown, but might be explained as a compensation for incorrect fitting of the intensity distribution and its complicated effect on the overall dynamics.  Removing the need to fit for intensity inhomogeneity would greatly simplify the QT protocol,  as we consider in future generations of our protocol, discussed in the next section. 

\begin{figure}[t]
\begin{center}
\includegraphics[width=8cm,clip]{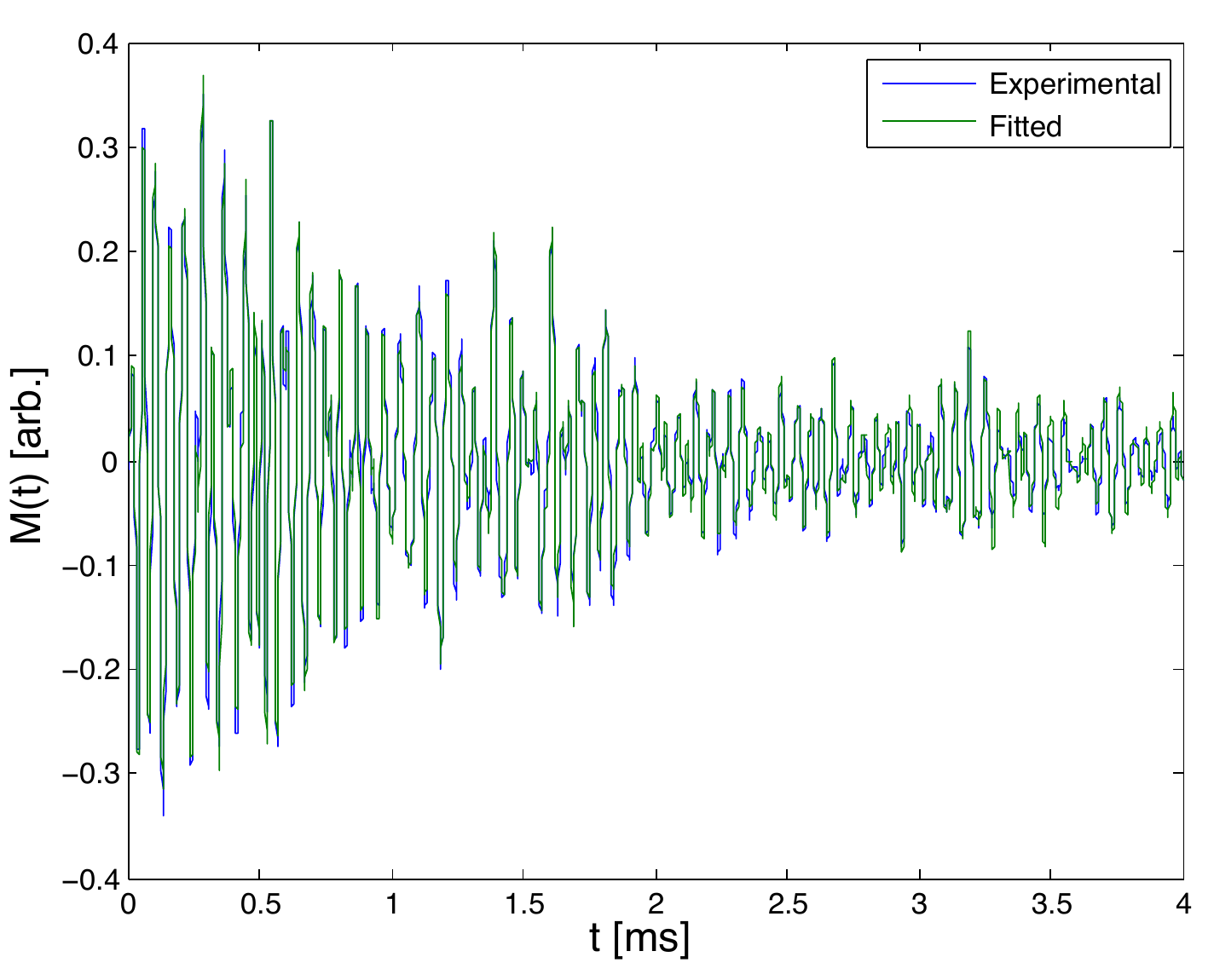}
\caption{(Color online) Comparison between the actual reconstruction signal (blue) and fitted by our procedure (green) for an initially prepared spin coherent state along $y$.}
\label{F:ReconstructionSignal}
\end{center}
\end{figure}

\begin{figure}[t]
\begin{center}
\includegraphics[width=8cm,clip]{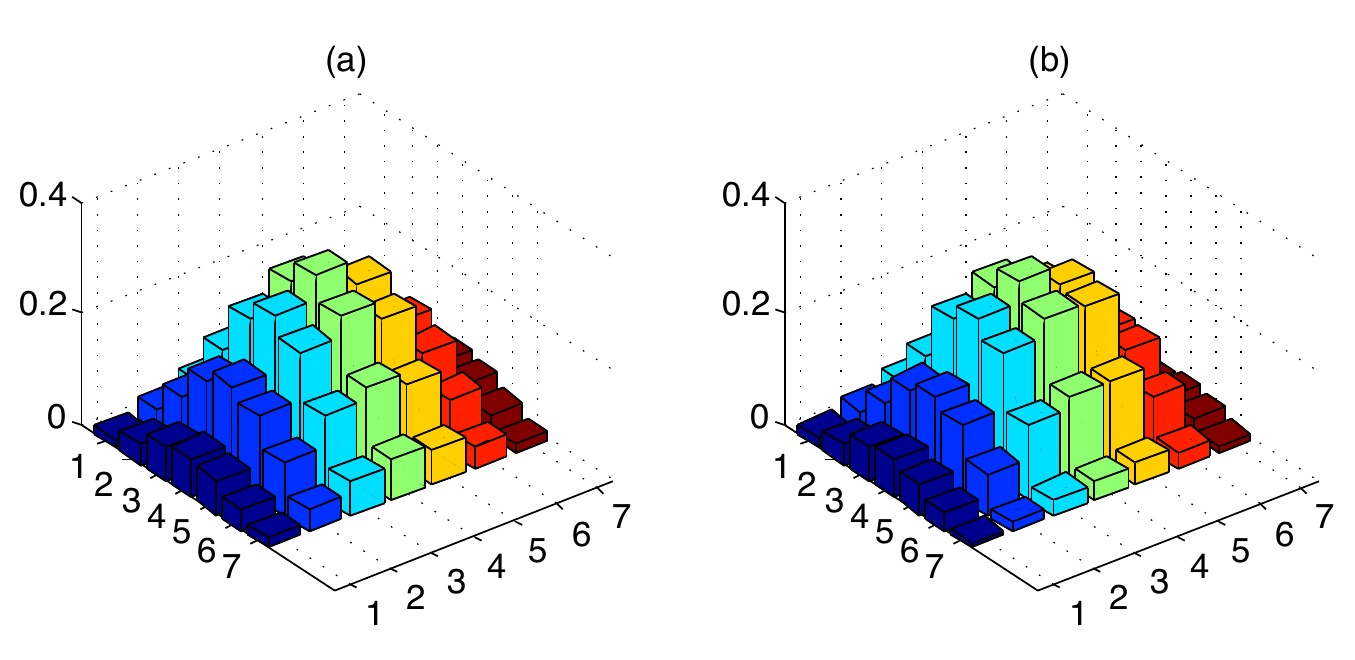}
\caption{(Color online) Comparison between the actual initial state (a) and the state reconstructed by our procedure (b). The absolute values of each element of the density matrix are plotted as a function of its index. The fidelity achieved by our procedure in this case was $\sim 0.95$.}
\label{F:Barplotls}
\end{center}
\end{figure}

Figure \ref{F:ReconstructionSignal} shows the fitted and the experimental reconstruction signal of a simple example: a spin coherent state along $y$. Once this fit is done, Eqs. (\ref{eq:convexprogram}) and (\ref{eq:convexprogramconstraint}) are used to find the closest positive-semidefinite estimate of the initial state. Figure \ref{F:Barplotls} shows a bar plot of the density matrix elements of the initial state and those of the reconstructed one. Qualitatively, we see how similar these two plots are and quantitatively we calculate the fidelity, Eq. (\ref{eq:fidelity}), to be about 0.95.  To compare this with the average expected behavior, including much more complex states produced in optimal control, we simulate measurement records based on our master equation, added noises, and signal processing.  For 1000 initial states randomly sampled using the Hilbert-Schmidt measure, we estimate our reconstruction fidelity to be about 0.998, limited only by the noise in the probe.  We attribute the lower fidelity of the experimental example above as arising from limitations in our ability to determine (estimate) all the parameters present in the actual system dynamics. Relaxing some of these limitations will increase the fidelities we can reach in experimental situations.  In the next section we describe our next-generation protocol that addresses some of these limitations while simultaneously allowing for reconstruction of the full 16-dimensional Hilbert space associated with the electronic ground-states.  Realization of this protocol would represent a substantial advance in complex quantum state reconstruction.  

\subsection{Reconstructing the entire hyperfine-manifold via rf and microwave control}\label{sec:CsRFUW}
	
In this section, we develop the tools necessary for reconstructing the entire 16-dimensional hyperfine ground state manifold ($F=3 \oplus F=4$) of cesium. To achieve this we make use of the type of control developed in \cite{merkel08}, which employs microwave ($\mu$w) and radio-frequency (RF) modulated external magnetic fields to drive the atoms. The RF-fields drive rotations on the $F=3$ and $F=4$ manifolds, whereas the microwave fields drive a resonant transition between two Zeeman levels in $F=3$ and $F=4$, thus making the system fully controllable.  Since the light-shift interaction is not necessary for controllability, in principle, the control can be achieved in a decoherence free way \cite{merkel08}. However, in practice, since a laser is used to measure the system, some decoherence will be present that must include in our model. 

The fundamental Hamiltonian, Eq. (\ref{eq:controlH}), can be written in this case $H(t)=H_0+H_{RF}+H_{\mu w}+H_{LS}$. The free Hamiltonian, $H_0$, includes the hyperfine interaction and the Zeeman shift produced by the bias magnetic field interaction, which is necessary to define the quantization axis and to resolve microwave-induced transitions of a pair of magnetic sublevels. Keeping terms to quadratic order in the field strength,
\be
\begin{split}
H_0=&\frac{\omega_{HF}}{2}\left(1+\frac{x^2}{2}\right)(P_4-P_3)+\Omega_0(F_z^{(4)}-g_r F_z^{(3)})\\
        &-\alpha((F_z^{(4)})^2-(F_z^{(3)})^2),
\end{split}
\label{eq:H0}
\ee
where $\omega_{HF}/2\pi \approx ~ 9.19$  GHz is the the hyperfine splitting for cesium, $P_F$, is the projector operator onto the $F=3,4$ manifolds, $\Omega_0=g_4\mu_BB_0$ is the Larmor frequency produced by the bias field $B_0$ in the $z$ direction, $x\equiv(g_e-g_I)\mu_BB_0/\omega_{HF}$ with $g_I\approx -0.0004$ and $g_e\approx 2.0023$ being the nuclear and electronic g-factors, respectively \cite{steck09}. The Land\'e g-factors for each manifold $F$ have opposite signs and are given by $g_3\approx 0.2499$ and $g_4\approx -0.2507$ and its ratio $g_r=|g_3/g_4|\approx 1.0032$ is the relative g-factor between the $F=3$ and $F=4$ manifolds. The last term in Eq. (\ref{eq:H0}) is the quadratic Zeeman shift, where $\alpha=x^2 \omega_{HF}/(2I+1)^2$. 

The RF-control Hamiltonian, $H_{RF}$, is produced by a RF-magnetic field that oscillates in the $x$ and $y$ directions
\be
\begin{split}
H_{RF}&=\Omega_x(t)\cos{(\omega_{RF}t-\phi_x(t))}(F_x^{(4)}-g_r F_x^{(3)})\\
             &+\Omega_y(t)\cos{(\omega_{RF}t-\phi_y(t))}(F_y^{(4)}-g_r F_y^{(3)}),
\end{split}
\label{eq:Hrf}
\ee
where the Larmor frequency is defined as $\Omega_{i}(t)=g_4\mu_{B}B_{i}(t)$, for $i=x,y$, $\phi_x(t)$, and $\phi_y(t)$ are the control phases of the magnetic field in the $x$ and $y$ directions respectively and $\omega_{RF}$ is the frequency at which the RF fields are modulated. This Hamiltonian allows for independent SU(2) rotations within the $F=3$ and $F=4$ manifolds.

For control via application of microwave radiation, we consider a purely $\sigma_+$-polarized field.  While in practice the polarization of microwaves is not well controlled at the position of the atoms, this is not critical, since ultimately we will drive a selected two-level transition through its unique resonance frequency.  Nonetheless, we allow for the effect of off-resonant ac-Stark shifts caused by microwaves and choose one polarization to analyze, for simplicity.  Under this assumption, the microwave control Hamiltonian is
\be
\begin{split}
H_{\mu w}&=\Omega_{\mu w}(t)\cos{(\omega_{\mu w}t-\phi_{\mu w}(t))}\times\\
		   &\sum_{m=-3}^{3}\braket{4,m+1}{3,m;1,1}\sigma_x^{(m)},
\end{split}
\label{eq:Huw}
\ee
where the bare microwave Rabi frequency is $\Omega_{\mu w}(t)=\mu_{B}B_{\mu w}(t)$, $\phi_{\mu w}(t)$ is its control phase, $\braket{4,m+1}{3,m;1,1}$ is the Clebsch-Gordan coefficient associated to the transition $|3,m\rangle\rightarrow|4,m+1\rangle$, and  $\sigma_x^{(m)}=\ket{4,m+1}\bra{3,m}+\ket{3,m}\bra{4,m+1}$. This Hamiltonian couples Zeeman levels in the two different manifolds, taking into account the resonant as well as the off-resonant transitions. 

With the control Hamiltonian in hand, we perform employ the rotating wave approximation (RWA) to eliminate the explicit time dependence. Following Appendix \ref{app:A}, the free Hamiltonian, Eq (\ref{eq:H0}), written in the rotating frame and after the RWA is given by Eq. (\ref{eq:AppH0R})
\be
\begin{split}
H'_{0}=&\left(\frac{3\Omega_0}{2}(1-g_r)+\frac{25}{2}\alpha\right)(P_4-P_3)\\
          &+\Omega_0(1-g_r)F_z^{(3)}-\alpha ((F_z^{(4)})^2-(F_z^{(3)})^2),
\end{split}
\label{eq:H0R}
\ee
where we consider the resonant case $\Delta_{RF}=\Delta_{\mu w}=0$. The RWA for the microwave Hamiltonian is straightforward since in general $\Omega_{\mu w} \ll \omega_{HF}$.  However, we must pay special attention to keep the correct off-resonant terms that lead to microwave-induced AC Stark shifts.  The resulting microwave Hamiltonian in the RWA is
\be
\begin{split}
H'_{\mu w}(t)&=\frac{\Omega_{\mu w}(t)}{2}\left[\cos{\phi_{\mu w}(t)}\sigma_x+\sin{\phi_{\mu w}(t)}\sigma_y\right]\\
		        &+\frac{\Omega_{\mu w}^2(t)}{8\omega_{RF}}\sum_{m\neq 3}\frac{|\braket{3,m;1,1}{4,m+1}|^2}{3-m}\sigma_z^{(m)},
\end{split}
\label{eq:Hmicrowave}
\ee
where $\sigma_z^{(m)}=\ket{3,m}\bra{3,m}-\ket{4,m+1}\bra{4,m+1}$, and $\sigma_x$ and $\sigma_y$ are given in Appendix \ref{app:A}.

The case of the RF-Hamiltonian is much more complex.  In order to maintain rapid control, the RF-Larmor rotation frequencies must be sufficiently large.  However, if the bias field is not sufficiently large, the condition $\Omega_x$, $\Omega_y$ $\ll \omega_{RF}$ will not be fulfilled.  In this case, we must consider higher order corrections to the RF Hamiltonian in the RWA. To do this, we follow \cite{fox87}, and employ the method of averages.  The details of this procedure are discussed in Appendix \ref{app:A}.  The resulting RF-Hamiltonian, Eq. (\ref{eq:Hrfcomplete}), is
\begin{widetext}
\be
\begin{split}
H_{RF}(t)=&\frac{\Omega_x(t)}{2}\left[\cos{\phi_x(t)}\left(F_x^{(4)}-\frac{(1+g_r)g_r}{2}F_x^{(3)}\right)-\sin{\phi_x(t)}\left(F_y^{(4)}+\frac{(3-g_r)g_r}{2}F_y^{(3)}\right)\right]+\\
&\frac{\Omega_y(t)}{2}\left[\cos{\phi_y(t)}\left(F_y^{(4)}-\frac{(1+g_r)g_r}{2}F_y^{(3)}\right)+\sin{\phi_y(t)}\left(F_x^{(4)}+\frac{(3-g_r)g_r}{2}F_x^{(3)}\right)\right]+\\
&\sum_{i=x,y}\frac{\Omega_i^2(t) }{16\omega_{RF}}\left(1-2\cos{2\phi_i(t)}\right) \left( F_z^{(4)} - g_r^2 F_z^{(3)} \right)+ \frac{\Omega_x(t)\Omega_y(t)}{8\omega_{RF}} \sin{\left(\phi_x(t)-\phi_y(t)\right)} \left( F_z^{(4)} + g_r^2 F_z^{(3)} \right)	
\end{split}
\label{eq:HradioF}
\ee
\end{widetext}
where, as before, we consider the resonant case and set $\omega_{RF}=\Omega_{0}$.
This differs from the Hamiltonian given in \cite{merkel08} in two ways.  We account for the relative magnitudes of the $g$-factors in the upper and lower manifolds due to the small nuclear magneton, $g_r \neq 1$.  Additionally we maintain terms of order $\Omega_{i}^2/\omega_{RF}$, $i=x,y$, which lead to Bloch-Seigert-like shifts and extra corrections due to counter-rotating terms.

We must also transform the light-shift Hamiltonian, Eq. (\ref{eq:effhamiltonianls}), to the rotating frame.  The RWA is an excellent approximation in this case since generally we will have parameters such that the Zeeman splitting is much larger than the rate of coherent coupling induced by the light shift, $\Omega_0 \gg \text{Re}(\beta^{(2)}) \gamma_{sc}$.  Making the appropriate unitary transformation and averaging over the rapidly varying terms, as described in Appendix \ref{app:A}, we obtain the effective light-shift Hamiltonian in each manifold $F$, 
\be
H'^{LS}_{{\rm eff},F}=\gamma_{sc} \left[\left( \beta_F^{(0)} + \beta_F^{(2)} \frac{F(F+1)}{6}  \right) I_F - \frac{\beta_F^{(2)}}{2} F_z^2\right].
\ee
Note, under the RWA, the light-shift does not drive coherences between magnetic sublevels defined by the quantization axis.  Such coherent couplings are no longer resonant in the presence of a strong bias field.  Also note that in considering dynamics over the full hyperfine manifold, we must retain the real part of the scalar light shift, since generally $\beta_3^{(0)} \ne \beta_4^{(0)}$.  The scalar contribution to the light shift thus drives coherences between the $F=3$ and $F=4$ manifolds.  

We now have all the ingredients to proceed with our simulations. The system evolves according to the master equation, Eq. (\ref{eq:csmaster}), with the effective Hamiltonian expressed in the RWA and the control Hamiltonian described above. In addition, special care must be taken when considering the full master equation. All operators, including the Lindblad jump operators, must be written in the rotating frame and the RWA should be applied accordingly. This is essential in order to account for spontaneous emission processes that become distinguishable once the energy degeneracy is broken by the shift produced by bias field. The RWA in the master equation is achieved by explicitly calculating  the transformation $U^{\dagger}(t)W_q^{F_bF_a}U(t)$, and averaging the superoperator map over the rapid oscillations.

\begin{figure}[t]
\begin{center}
\includegraphics[width=8.7cm,clip]{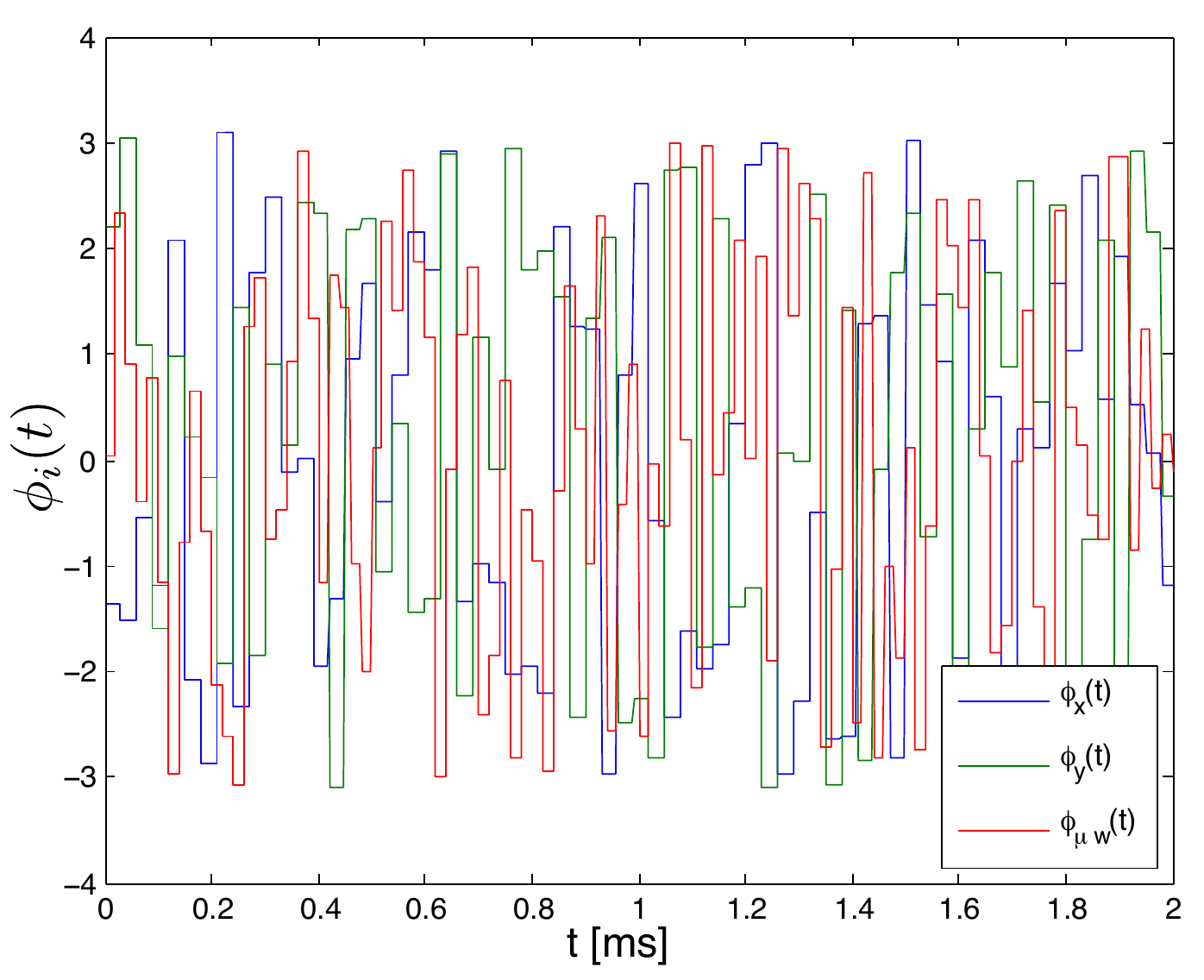}
\caption{(Color online) Randomly sampled control waveforms that determine the phases of the applied RF and microwave magnetic fields as given in Eqs. (\ref{eq:Hmicrowave},\ref{eq:HradioF}). These waveforms produce high fidelity estimates with reasonable stability.}
\label{F:RFuwcontrols}
\end{center}
\end{figure}

Following \cite{merkel08}, full controllability of the system can be achieved by keeping the RF-Larmor and $\mu$w-Rabi frequencies, $\Omega_x$, $\Omega_y$, and $\Omega_{\mu w}$, constant in time, while varying the control phases $\phi_x(t)$, $\phi_y(t)$, and $\phi_{\mu w}(t)$. Due to the size of the Hilbert space, finding a set of control waveforms is a very challenging task.  Optimizing the entropy of the Gaussian probability distribution, as mentioned in Sec. \ref{sec:CsLightShift}, is generally an intractable problem. Instead, we choose the control waveforms as piecewise random functions. We have found in our numerics that this is sufficient to generate an informationally complete measurement record. For the RF ($\mu$w) waveforms, the phase is chosen uniformly between $-\pi$ and $\pi$ and kept constant over intervals of 30$\mu$s (20$\mu$s). After a total time of 2 ms we ensure that we have sufficient information in the measurement for QT. Fig. \ref{F:RFuwcontrols} shows an example of the control phases used in this work. In general, almost all the control waveforms designed in this random way will produce informationally complete measurement records for most initial states. However, numerical stability during the use of Eq. (\ref{eq:LSsolution}) becomes an issue for certain waveforms. We thus choose a set of waveforms that produce the most stable results by repeating the design procedure several times.

To complete our protocol, we must choose the detuning $\Delta_c$ of the probe.  In the previous section, this detuning was chosen to maximize the nonlinear light shift relative to photon scattering.  In the current context, we have much more flexibility, since full control of the Hilbert space can be achieved without the light shift.  There are, however, many technical considerations that inform the choice of detuning.  Firstly, the measurement strength is proportional to $\gamma_{sc}$, so we can never make the measurement record free of decoherence by detuning further off resonance \cite{smith03}.  Moreover, at very large detunings, in order to maintain a reasonably large $\gamma_{sc}$, we would need a large probe intensity for which shot-noise-limited detection is difficult.  For these reasons, the light-shift-driven dynamics must be included in the analysis.  An additional technical issue is the effect of inhomogeneity in the light intensity across the ensemble.  Indeed, the difficulty in estimating the distribution of intensities caused substantial complexity in the reconstruction algorithm, as discussed in Sec. \ref{sec:CsLightShift}, and ultimately limited the fidelity of the protocol.  Mitigating this effect would greatly improve the performance.  

From Eq. (\ref{eq:effhamiltonianls}), we see that the scalar, $\text{Re}(\beta^{(0)})$ and tensor, $\text{Re}(\beta^{(2)})$, light shift components are responsible for the introduction of inhomogeneity in the problem. While there is no choice of detuning that makes both terms exactly zero, the scalar light shift will cause the largest problems, and our goal is to cancel it.  Putting all of these considerations together, we choose a relatively small detuning where the measurement strength can be large at low intensity.  For such a detuning, only one $F$ manifold is effectively coupled to the light, and the other is so far from resonance that its coupling is negligible.  We chose this to be $F=3$, and detune of the D1 line, $(6s_{1/2})F=3$ to $(6p_{1/2})F'=3$.  Detuned within the hyperfine splitting of the excited state, we can find a ``magic wavelength'' at which $\text{Re}(\beta^{(0)}_3)=0$.  Using the tensor coefficients given in \cite{deutsch09}, we find the magic detuning on the D1 line is $\Delta_{c}/2\pi = 291.89$ MHz.  The residual light shift is due to the tensor term, proportional to $\text{Re}(\beta^{(2)}_3)=1.35$, which together with microwave and RF fields, drives the spin dynamics during the course of the measurement.

\begin{figure}[t]
\begin{center}
\includegraphics[width=8.7cm,clip]{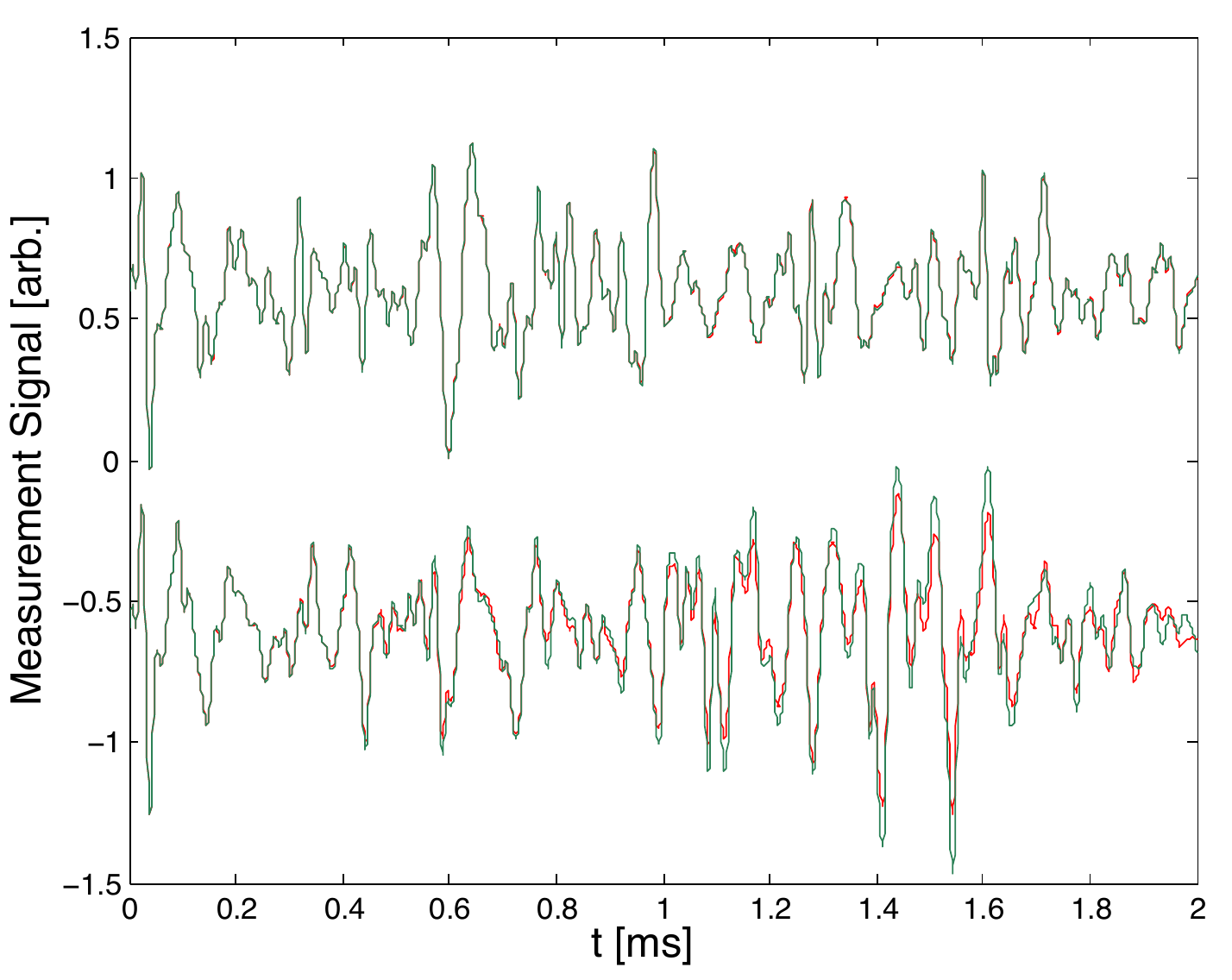}
\caption{(Color online) Simulated measurement signal for a random pure state chosen from the Haar measure for different choices of detuning of the laser probe. (Top) $\Delta_{c}/2\pi=291.89$MHz, the ``magic wavelength'' at which the scalar light shift is set to zero for $F=3$. The red line is an averaged signal over a Gaussian distribution of intensity. The green line is the signal that a system would produce if it evolves under the nominal value of intensity. (Bottom) $\Delta_{c}/2\pi=642.78$MHz, the detuning used in previous experiments. The red line is an averaged signal over a Gaussian distribution of intensity. The green line is the signal that a system would produce if it evolves under the nominal value of intensity.  The signal is more robust at the ``magic wavelength".}
\label{F:signalcomparison}
\end{center}
\end{figure}

To study the performance of our protocol, we performed numerical simulations of the expected measurement signal for a random pure state sampled from the Haar measure.  We choose the following parameters: $\Omega_{x}/2\pi=\Omega_{y}/2\pi=15$ kHz, $\Omega_{\mu w}/2\pi=33$ kHz, $\Omega_0/2\pi=1.0$ MHz, and a characteristic photon scattering rate $\gamma_{sc}/2\pi=410.7$ Hz.  Furthermore, we add Gaussian white noise to the signal so that the signal-to-noise ratio is 100.  As a first check, confirm that our choice of detuning makes our system more robust to light-shift inhomogeneity, by adding Gaussian fluctuations in the intensity across the ensemble, and then averaging the result.  For example, if we choose the same detuning as used in Sec. \ref{sec:CsLightShift}, $\Delta_c/2\pi=642.78$ MHz, $\beta^{(0)}_3\ne 0$, and we see that  the average signal leads to fidelities $\le 0.80$, whereas a similar simulation with the optimized detuning, $\Delta_c/2\pi=291.89$ MHz produces fidelities $\ge 0.90$.  Fig. \ref{F:signalcomparison} shows qualitatively a comparison between the simulated measurement records, averaged over a Gaussian distribution of intensities for these two detunings  and the simulated signals with a fixed, nominal value of intensity.  It is clear that the optimized detuning produces much better results, making both averaged and nominal signals look very similar. This will simplify our procedure for estimating the intensity distribution seen by the atomic ensemble. A fit to a Gaussian distribution will be sufficient to capture the effects of the inhomogeneous light shift.

A number of calibration errors are possible in this system. We anticipate the need of fitting similar parameters to the ones in the last section, i.e., the RF and $\mu$w magnetic field amplitudes as well as the origin of time. Least squares techniques, similar as the ones discussed in detail in Sec. \ref{sec:CsLightShift}, should suffice to achieve good accuracy parameter estimation and high-fidelity reconstructions. Finally, we choose to measure Faraday rotation instead of birefringence of the probe, which further simplifies the protocol since we no longer must fit for the initial measurement operator.  Our simulations below show that this choice leads to high fidelity QT.  Thus, for the simulations shown in this section, we have chosen our initial observable to be the Faraday operator $\cO_0= \vec{\mathcal{O}}\cdot\vec{e}_3$ as given in Eq. (\ref{eq:FaradayObservable}).

We have run several simulations to test the performance and efficiency of this protocol. We numerically generated a measurement record for different initial states and different noise realizations and amplitudes according to Eq. (\ref{eq:measurement}). Then a Bessel bandpass filter from 6 to 80 kHz was applied to the simulated measurement record in order to limit the noise in frequency components that are not present in the measurement. The same filter is applied the Heisenberg picture observable $\cO(t)$ to account for all dynamical effects the signal undergoes. Once this is done, Eq. (\ref{eq:LSsolution}) is used to find the unconstrained maximum likelihood estimate of the initial state and the convex program Eq. (\ref{eq:convexprogram}) is solved to find the physical density matrix that best represent the measured data. In order to quantify the performance of our method, we calculate the the fidelity between the initial and estimated state,  Eq. (\ref{eq:fidelity}).

\begin{figure}[t]
\begin{center}
\includegraphics[width=8.7cm,clip]{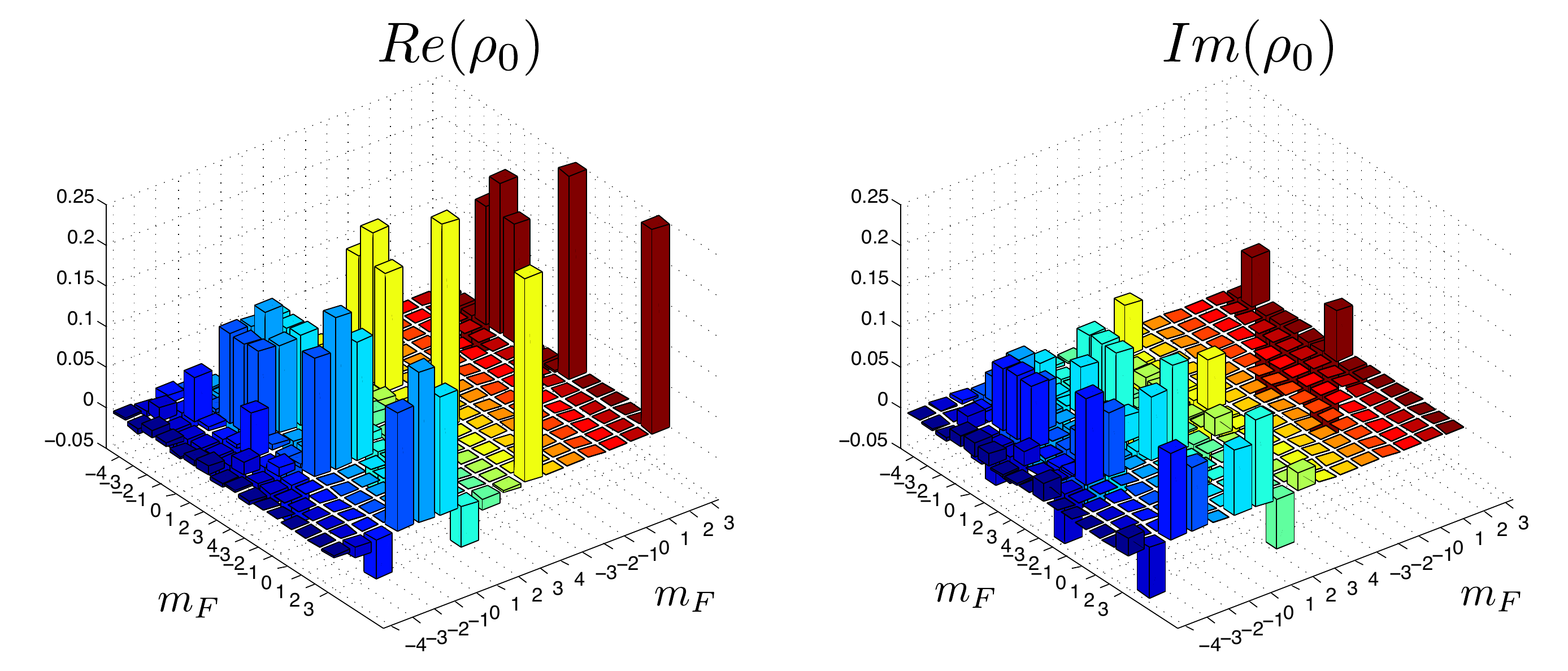}
\caption{(Color online) Real and imaginary parts of the elements of $\rho_0$, the initial state used in the simulations. This is a non-trivial state used for illustration, $\ket{\psi}=\left( \ket{\psi^{(4)}_{sq}}+\ket{\psi^{(3)}_{cat}} \right) / \sqrt{2}$, consisting of an equal superposition of a spin squeezed state in the $F=4$ manifold, $\ket{\psi^{(4)}_{sq.}}=\exp\left\{-i0.5F_z^2\right\} \ket{F=4,m_x=4}$ and a ``cat state"  $\ket{\psi^{(3)}_{cat}}=\left(\ket{F=3,m_z=3}+\ket{F=3,m_z=-3}\right)/\sqrt{2}$, in the $F=3$ manifold.}
\label{F:InitialState}
\end{center}
\end{figure}

\begin{figure}[t]
\begin{center}
\includegraphics[width=8.7cm,clip]{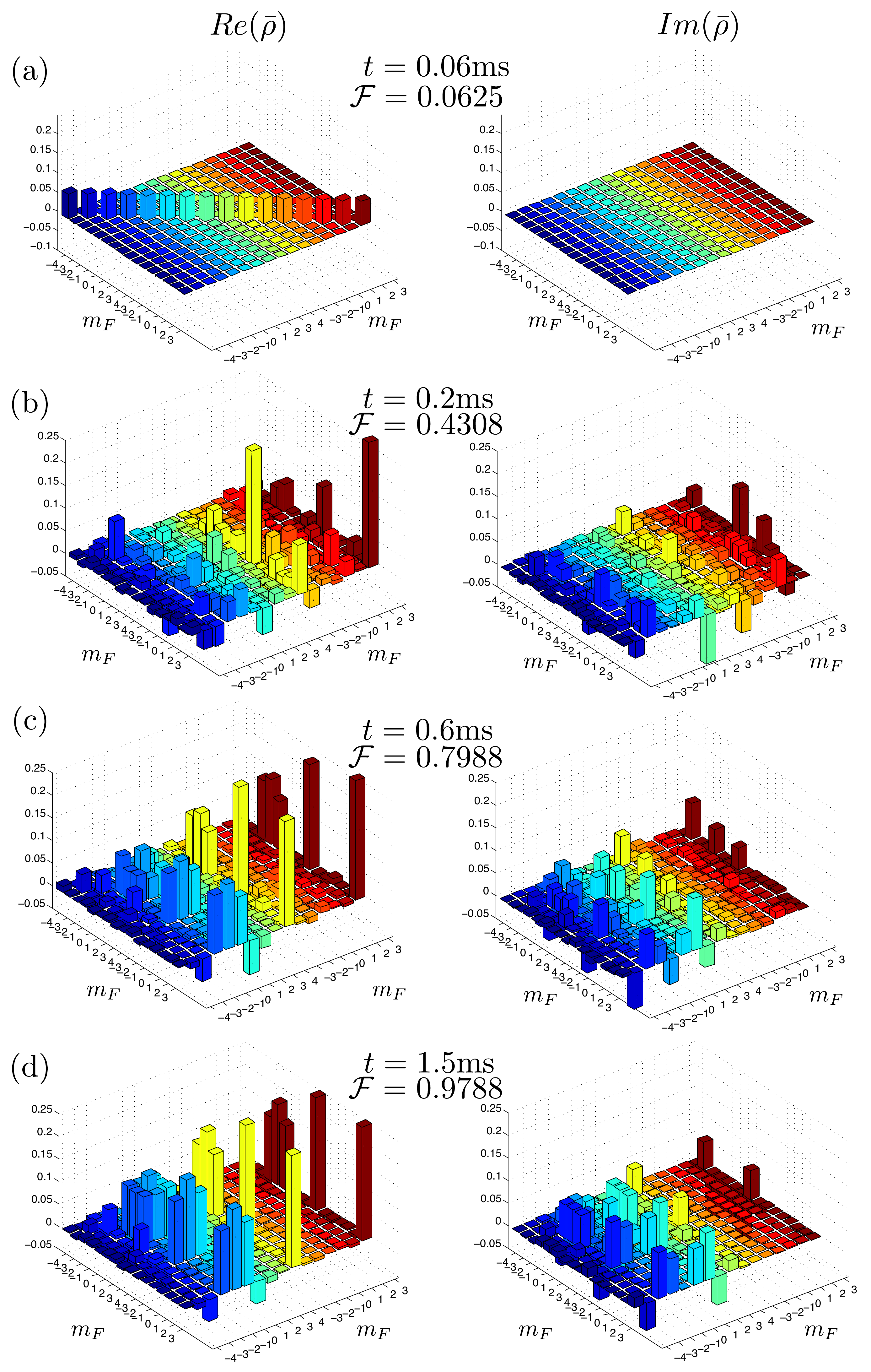}
\caption{(Color online) Real and imaginary parts of the estimated initial state $\bar{\rho}$. (a) After 60$\mu$s not enough information has been collected, thus there is no convergence of our algorithm, and the maximally mixed state is guessed. (b) and (c) More information is acquired as time passes and higher fidelities are obtained. (d) A high fidelity estimate is obtained after 1.5ms of simulation. Slightly higher fidelities are seen for longer simulation times.}
\label{F:Frames}
\end{center}
\end{figure}

As an example, we simulate the reconstruction the nontrivial state, $\ket{\psi}=\left( \ket{\psi^{(4)}_{sq}}+\ket{\psi^{(3)}_{cat}} \right) / \sqrt{2}$, shown in Fig. \ref{F:InitialState}, consisting of an equal superposition of a spin squeezed state in the $F=4$ manifold, $\ket{\psi^{(4)}_{sq}}=\exp\left\{-i0.5F_z^2\right\}\ket{F=4,m_x=4}$ and a ``cat state"  in the $F=3$ manifold $\ket{\psi^{(3)}_{cat}}=\left(\ket{F=3,m_z=3}+\ket{F=3,m_z=-3}\right)/\sqrt{2}$. In Fig. \ref{F:Frames}, we show how our procedure converges as a function of time to an estimate of the initial state with high fidelity. Within the few first microseconds of the simulation there is little information, and the protocol returns the maximally mixed state as the estimation of $\rho_0$. However, as the time passes, more information about the informationally complete set observable is acquired and the protocol makes better guesses of the initial state.  For a SNR of 100 simulated here, within 1 ms, a fidelity of $>0.97$ is achieved.

\begin{figure}[t]
\begin{center}
\includegraphics[width=8cm,clip]{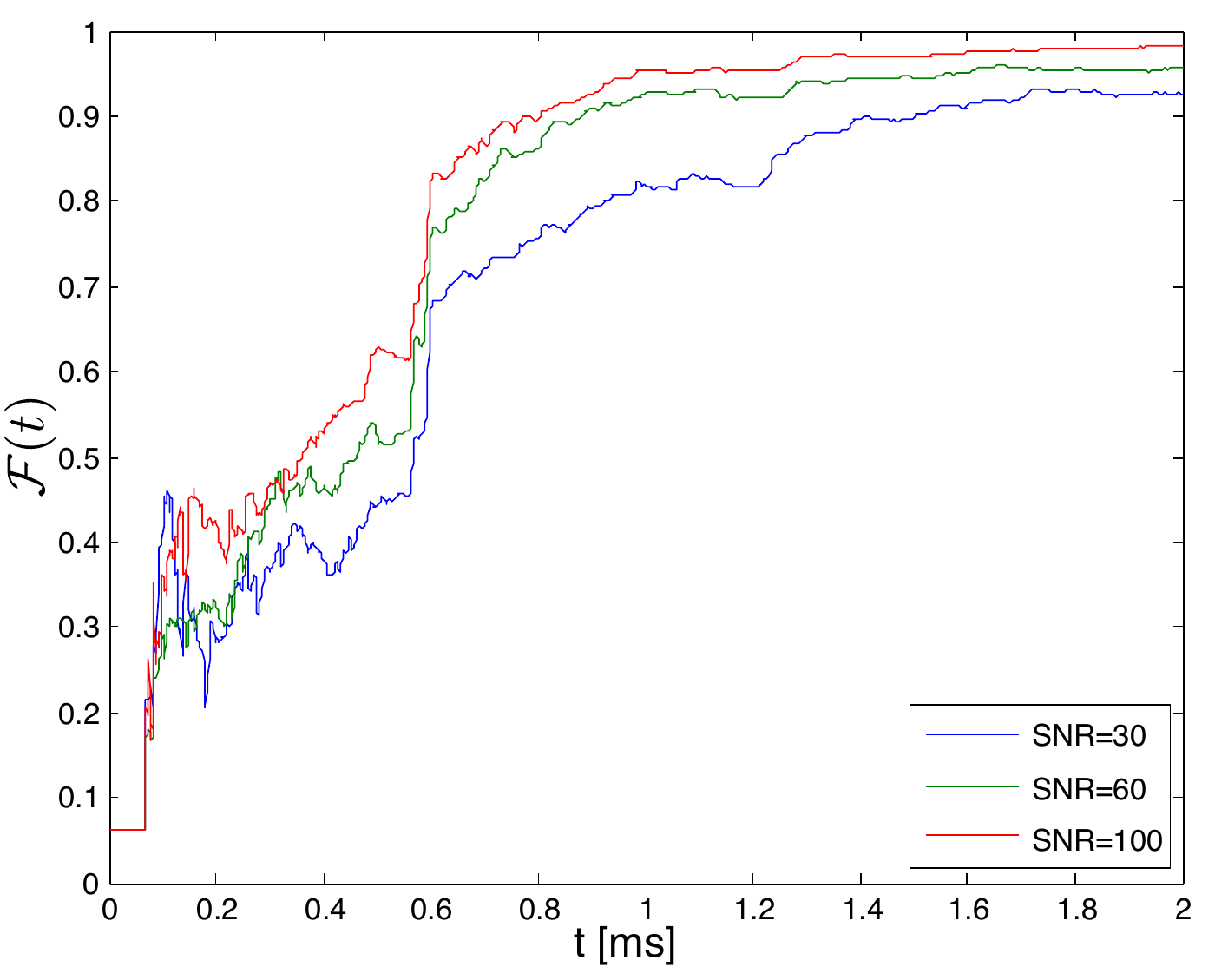}
\caption{(Color online) Fidelity, Eq. (\ref{eq:fidelity}), of the reconstructed state respect to the actual, random, pure state as a function of time. A jump in the fidelity is seen at $t \approx 0.8$ms, before an informationally complete measurement set is obtained, due to the positivity constraint.  Shown is the performance of the QT protocol under different signal-to-noise ratios  (SNR).  A fidelity of $\sim 0.977$ is seen for a SNR of 100.}
\label{F:FidelityTime}
\end{center}
\end{figure}

In Fig. \ref{F:FidelityTime}, we show the fidelity between the estimated state and the initial state for a random pure state in the 16 dimensional Hilbert space as a function of time for different signal-to-noise ratios. Although this plot shows the performance of the reconstruction protocol for a particular state, the same behavior is seen for most random states sampled from the Haar measure. In fact, for 200 of such random pure states, we achieved an average fidelity $\approx 0.977$ with standard deviation of $\approx 0.006$ after 2ms and an SNR of 100.

\section{Summary and outlook}
In this work we have presented a comprehensive review of a protocol to perform fast, robust, high-fidelity quantum tomography (QT) based on continuous measurement of an informationally complete set of observables.  This procedure is applicable when one has access to a large ensemble of identically prepared systems in a product state, collectively coupled to a probe field.  For weak measurement back-action, the probability distribution of parameters that define the density matrix, conditioned on the measurement record, is Gaussian, and the problem maps onto one of classical stochastic parameter estimation.  With sufficient signal-to-noise, the density matrix can be found from a single measurement record using a maximum-likelihood estimate.  A physical density matrix, with positive eigenvalues, is then found using a convex optimization algorithm that searches for the closest positive state to the unconstrained maximum-likelihood prediction.

We have applied this protocol to the problem of QT of hyperfine spins in cold atomic ensembles.  A key component of our procedure is to drive the system with well-chosen control fields so as to generate an informationally complete set of observables over the course of the measurement record.  We have presented here two methods for achieving this, one based on combinations of magnetic-generated Larmor precession and a nonlinear spin rotation generated by the light-shift, and another approach based on combinations of microwave and radio-frequency-driven spin rotations.  The former allows us to reconstruct the density matrix associated with one hyperfine manifold $F$, as has been demonstrated in experiments on the 7-dimensional $F=3$ manifold of Cs atoms \cite{smith06}.  The latter is more ambitious, allowing us to reconstruct the entire electronic ground state subspace (16 Hilbert space dimensions for Cs).   

Our protocols rests on the assumption that once the measurement record is obtained, we can invert its history to estimate the initial quantum state of the ensemble.  It is thus essential to accurately model the atomic dynamics and measurement of the observables, including known sources of imperfections.  We have given a detailed discussion of the master equation governing the atomic dynamics and a measurement model based on polarization spectroscopy.  With these in hand, we showed how to simulate the measurement record, with emphasis on limitations, challenges, and the steps needed to make our protocol reliable and stable.  For the case of light-shift control, we tested our state-of-the-art protocol with experimental data and found that we could efficiently reconstruct the 7-dimensional quantum state with a fidelity 0.95, limited primarily by the difficulty in accounting for the effects of the inhomogeneous intensity.    Removing this constraint, our simulations show that we should obtain a fidelity $>$0.99 with a SNR of 100.  In the case of full control on the 16-dim spin system via RF and microwave driving, we simulated noisy measurement records and used these as inputs to our reconstruction algorithm.  We found that we can rapidly achieve average fidelity $>$0.97 at the same SNR for a measurement time of only 2 ms.  These initial results bode well for high-fidelity reconstruction of a quantum state in such a large Hilbert space.  Experimental studies are underway.  With such a tool, we can explore the implementation of qudit unitary transformations for quantum information processing \cite{merkel09}, and nontrivial dynamics as seen in quantum chaos \cite{chaudhury09}. 

Our approach to quantum tomography could be improved in a number of ways.  While we have found that random waveforms are sufficient for generating informationally complete measurement records, the nature of optimal waveforms (in time and/or average fidelity) remains open.  Additionally, as the protocol has analogies with classical stochastic estimation, we see potential for improving the reliability and stability of the reconstruction procedure by employing the data processing tools such as Kalman filters \cite{maybeck79} and other methods of estimation theory, which we plan to explore in future studies.

\acknowledgments 
We thank Aaron Smith, Brian Anderson, and Robin Blume-Kohout for helpful discussions.  This work was supported by NSF Grant PHY-0903692 and the CQuIC NSF grant PHY-0903953.

\appendix
\section{RWA corrections}\label{app:A}
This appendix describes in detail the derivation of the Hamiltonian that governs the dynamics hyperfine manifold driven by RF-and microwave magnetic fields as well as light shift interactions.  In general, for applicable parameters, we must go beyond the usual linear Zeeman effect and first order rotating wave approximation, which substantially complicates the Hamiltonian beyond that presented in \cite{merkel08}.  To begin we transform to a frame that is rotating at the frequency of the control fields, according to the unitary transformation $U(t)=U_{RF}U_{\mu w}$, where 
\begin{subequations}
\begin{align}
U_{RF}=\exp{[-i\omega_{RF}t(F_z^{(4)}-F_z^{(3)})]}\\
U_{\mu w}=\exp{[-i\frac{\theta t}{2}(P_4-P_3)]}
\end{align}
\end{subequations}
with $\theta=\omega_{\mu w}-(m_4+m_3)\omega_{RF}$, where $m_4$ and $m_3$ label two Zeeman levels corresponding to the $F=4$ and $F=3$ manifolds respectively. It then follows from  Eq. (\ref{eq:H0}), 
\begin{widetext}
\be
\begin{split}
H'_{0}&=U^{\dagger}(t)H_0U(t)-iU^{\dagger}\frac{dU}{dt}\\
           &= \left(\frac{3\Omega_0}{2}(1-g_r)+\frac{25}{2}\alpha+\frac{1}{2}(7\Delta_{RF}-\Delta_{\mu w})\right )(P_4-P_3)-\Delta_{RF} F_z^{(4)} +\left(  \Delta_{RF} +\Omega_0(1-g_r) \right) F_z^{(3)}-\alpha ((F_z^{(4)})^2-(F_z^{(3)})^2),
\end{split}
\label{eq:AppH0R}
\ee
\end{widetext}
where we have chosen $m_4=4$, $m_3=3$, $\Delta_{RF}=\omega_{RF}-\Omega_0$, $\Delta_{\mu w}=\omega_{\mu w}-\omega_0$, with $\omega_0=\omega_{HF}+(4+3g_r)\Omega_0+7\alpha$ being the on-resonant transition for the two-level system formed by the stretched states $|3,3\rangle$ and $|4,4\rangle$. Although our goal is to be as close to resonance as possible ($\Delta_{RF}=\Delta_{\mu w}=0$), in practice we must also account for nonzero detunings that might result, e.g., from gradients in $\Omega_0$ across the ensemble.

Going to the rotating frame, the RF Hamiltonian in Eq. (\ref{eq:Hrf}) transforms to $H'_{RF}(t)=U^{\dagger}(t)H_{RF}(t)U(t)$, yielding
\begin{widetext}
\be
\begin{split}
H'_{RF}(t)&=\frac{\Omega_x(t)}{2}(\cos{(2\omega_{RF}t-\phi_x(t))}+\cos{(\phi_x(t)}))(F_x^{(4)}-g_r F_x^{(3)})-\frac{\Omega_x(t)}{2}(\sin{(2\omega_{RF}t-\phi_x(t))}+\sin{(\phi_x(t)}))(F_y^{(4)}+g_r F_y^{(3)})\\
                  &+\frac{\Omega_y(t)}{2}(\cos{(2\omega_{RF}t-\phi_y(t))}+\cos{(\phi_y(t)}))(F_y^{(4)}-g_r F_y^{(3)})+\frac{\Omega_y(t)}{2}(\sin{(2\omega_{RF}t-\phi_y(t))}+\sin{(\phi_y(t)}))(F_x^{(4)}+g_r F_x^{(3)})
\end{split}
\label{eq:HrfRot}
\ee
\end{widetext}
In the same manner, we can write the microwave Hamiltonian, Eq. (\ref{eq:Huw}), in the rotating frame $H'_{\mu w}(t)$
\begin{widetext}
\be
\begin{split}
H'_{\mu w}(t)&=\frac{\Omega_{\mu w}(t)}{2}(\cos{(2\omega_{\mu w}t-\phi_{\mu w}(t))}+\cos{(\phi_{\mu w}(t)}))\sigma_x+\frac{\Omega_{\mu w}(t)}{2}(\sin{(2\omega_{\mu w}t-\phi_{\mu w}(t))}+\sin{(\phi_{\mu w}(t)}))\sigma_y\\
                  &+\frac{\Omega_{\mu w}(t)}{2}\sum_{m\neq 3}\braket{3,m;1,1}{4,m+1}(\cos{(2\omega_{\mu w}t+2(m-3)\omega_{RF}t-\phi_{\mu w}(t))}+\cos{(2(m-3)\omega_{RF}t+\phi_{\mu w}(t)}))\sigma_x^{(m)}\\
                  &+\frac{\Omega_{\mu w}(t)}{2}\sum_{m\neq 3}\braket{3,m;1,1}{4,m+1}(\sin{(2\omega_{\mu w}t+2(m-3)\omega_{RF}t-\phi_{\mu w}(t))}+\sin{(2(m-3)\omega_{RF}t+\phi_{\mu w}(t)}))\sigma_y^{(m)}
\end{split}
\label{eq:HuwRot}
\ee
\end{widetext}
where $\sigma_y^{(m)}=-i\ket{3,m}\bra{4,m+1}+i\ket{4,m+1}\bra{3,m}$. Note that we have explicitly separated the resonant terms from the off-resonant ones. The off-resonant interaction produces a AC-Zeeman shift of the magnetic levels that must be accounted for in the regime we consider in our simulations. We have defined for the resonant transition $\sigma_x=\sigma_x^{(3)}$ and $\sigma_x=\sigma_y^{(3)}$.  Finally, the effective light-shift Hamiltonian, given by Eq. (\ref{eq:lshamiltonian}) written in its irreducible tensor representation (as in Eq. (\ref{eq:effhamiltonianls})), can be expressed in the rotating frame as
\be
\begin{split}
H'^{LS}_{{\rm eff}}=&\gamma_{sc}\sum_F\left[\left(\beta_F^{(0)}+\beta_F^{(2)}\frac{F(F+1)}{6}\right)I_F\right.\\ 
                              &-\left.\frac{\beta_F^{(2)}}{2} (F_x^{(F)}\cos{(\omega_{RF}t)}+F_y^{(F)}\sin{(\omega_{RF}t)})^2\right],
\end{split}
\label{eq:LShamiltonianRot}
\ee
where $\beta_{F}^{(K)}$ is given in Eqs. (\ref{eq:betas}).

Given the Hamiltonian in the rotating frame, we proceed to apply the RWA.   Typically this is a straightforward task, equivalent to dropping the rapidly oscillating counter-rotating terms.  For the case RF Hamiltonian, since the Zeeman splitting induced by the basis magnetic field, $\Omega_0$ may not be much larger that the driving frequency, $\omega_{RF}$, a second order correction of the RWA is needed. In the remainder of this appendix, we follow \cite{fox87} and use the method of averages for ordinary differential equations, which we briefly review, to provide the required correction to the RWA.

Given a set of first order differential equations of the form $\frac{d{\bf x}}{dt}=\epsilon~ {\bf f}({\bf x},t,\epsilon)$, where ${\bf x}$ represents the state of the system, ${\bf f}({\bf x},t,\epsilon)$ is a periodic function with period $T$, and $\epsilon$ is a small parameter, we seek an approximate solution of the equivalent averaged system ${\bf y}$ under the transformation ${\bf x}={\bf y}+\epsilon~{\boldsymbol \omega}({\bf y},t,\epsilon)$, where ${\boldsymbol \omega}({\bf y},t,\epsilon)$ is also periodic with period $T$. The averaging theorem says that the equations of motion of the equivalent system are $d{\bf y}/dt=\epsilon~ {\bf \bar{f}}({\bf y})+\epsilon^2~{\bf f}_1({\bf y},t,\epsilon)+O(\epsilon^3)$, where ${\bf \bar{f}}({\bf y})=\frac{1}{T}\int_0^T{\bf f}({\bf y},t,\epsilon)dt$

Noting that when only the RF part of the Hamiltonian is present in the problem, the complete dynamics of the system can be described in the SU(2) group, and thus, all the dynamics of the system can be described by the Heisenberg equations of motion of $F_x$, $F_y$ and $F_z$. We carry out this calculation for the $F=4$ and $F=3$ manifolds separately since there is no coupling between them in the absence of the microwaves. Moreover, we assume a small enough bias field $B_0$ so that we can neglect the quadratic Zeeman shift introduced in Eq. (\ref{eq:H0});  for a very large bias the standard RWA is sufficient. For illustration, we discuss in detail the second order correction to the $F=4$ manifold RF Hamiltonian.

In this case, it is convenient to define the small parameter $\epsilon=\epsilon_0/\omega_{RF}$ where $\epsilon_0=\sqrt{\Omega_x^2(t)+\Omega_y^2(t)}$ to allow the RF Larmor frequencies to be different.  Turning the microwave Hamiltonian off ($\Omega_{\mu w}=0$) and neglecting the second order Zeeman shift, the Hamiltonian restricted to the $F=4$ manifold can be written
\be
\begin{split}
H^{(4)}(t)&=\left(\frac{3\Omega_0}{2}(1-g_r)+\frac{1}{2}(7\Delta_{RF}-\Delta_{\mu w})\right)I^{(4)}\\
	        &+\epsilon_0\frac{\chi(t)}{2}F_x^{(4)}+\epsilon_0\frac{\upsilon(t)}{2}F_y^{(4)}-\Delta_{RF}F_z^{(4)}
\end{split}
\label{eq:H4}
\ee
where 
\be
\begin{split}
\chi(t)&=\frac{\Omega_x(t)}{\epsilon_0}(\cos{(2\omega_{RF}t-\phi_x(t))}+\cos{(\phi_x(t))})\\
	  &+\frac{\Omega_y(t)}{\epsilon_0}(\sin{(2\omega_{RF}t-\phi_y(t))}+\sin{(\phi_y(t))}),
\end{split}
\ee
and
\be
\begin{split}
\upsilon(t)=&-\frac{\Omega_x(t)}{\epsilon_0}(\sin{(2\omega_{RF}t-\phi_x(t))}+\sin{(\phi_x(t))})\\
	  &+\frac{\Omega_y(t)}{\epsilon_0}(\cos{(2\omega_{RF}t-\phi_y(t))}+\cos{(\phi_y(t))}).
\end{split}
\ee

The Heisenberg equations of motion for the components of the total angular momentum can then be easily written
\begin{subequations}
\begin{align}
\frac{dF_x^{(4)}}{dt'}&=\epsilon\left(\frac{\upsilon(t')}{2}F_z^{(4)}+\frac{\Delta_{RF}}{\epsilon_0}F_y^{(4)}\right),\\
\frac{dF_y^{(4)}}{dt'}&=-\epsilon\left(\frac{\chi(t')}{2}F_z^{(4)}+\frac{\Delta_{RF}}{\epsilon_0}F_x^{(4)}\right),\\
\frac{dF_z^{(4)}}{dt'}&=\epsilon\left(\frac{\chi(t')}{2}F_y^{(4)}-\frac{\upsilon(t')}{2}F_x^{(4)}\right),
\end{align}
\end{subequations}
where we have scaled the time so that $t'=\omega_{RF}t$. This system of differential equations is in the form needed to apply the averaging theorem when we note that
\be
{\bf x}\rightarrow\left[\begin{array}{c} F_x^{(4)}\\F_y^{(4)}\\F_z^{(4)}\end{array}\right],~ {\bf f}({\bf x},t')\rightarrow\frac{1}{2}\left[\begin{array}{c} \upsilon(t')F_z^{(4)}+2\tilde{\Delta}F_y^{(4)}\\-\chi(t')F_z^{(4)}-2\tilde{\Delta}F_x^{(4)}\\ \chi(t')F_y^{(4)}-\upsilon(t')F_x^{(4)}   \end{array}\right],
\ee
where, for convenience, we have defined $\tilde{\Delta}=\Delta_{RF}/\epsilon_0$.
Transforming the original system to the averaged equivalent one, we have
\be
{\bf y}\rightarrow\left[\begin{array}{c} \bar{F}_x^{(4)}\\ \bar{F}_y^{(4)}\\ \bar{F}_z^{(4)}\end{array}\right],~ {\bf \bar{f}}({\bf y})\rightarrow\frac{1}{2}\left[\begin{array}{c} \bar{\upsilon}\bar{F}_z^{(4)}+2\tilde{\Delta}F_y^{(4)}\\-\bar{\chi}\bar{F}_z^{(4)}-2\tilde{\Delta}F_x^{(4)}\\ \bar{\chi}\bar{F}_y^{(4)}-\bar{\upsilon}\bar{F}_x^{(4)}   \end{array}\right]
\ee
where
\be
\bar{\chi}=\frac{\Omega_x(t)}{\epsilon_0}\cos{(\phi_x(t))}+\frac{\Omega_y(t)}{\epsilon_0}\sin{(\phi_y(t))},
\ee
and
\be
\bar{\upsilon}=-\frac{\Omega_x(t)}{\epsilon_0}\sin{(\phi_x(t))}+\frac{\Omega_y(t)}{\epsilon_0}\cos{(\phi_y(t))}.
\ee
The only terms that participate in the averaging process are the fast oscillating ones, while the slow varying terms are treated as constant.

We now proceed to calculate the function ${\boldsymbol \omega}({\bf y},t')=\int_0^{t'}({\bf f}({\bf y},t'')-{\bf\bar{f}}({\bf y}))dt''$ by again integrating only over the fast varying terms
\be
{\boldsymbol \omega}({\bf y},t')\rightarrow\frac{1}{4}\left[\begin{array}{c} \Upsilon(t')F_z^{(4)}\\-{\rm X}(t')F_z^{(4)}\\ {\rm X}(t')F_y^{(4)}-\Upsilon(t')F_x^{(4)}   \end{array}\right]
\ee
where 
\be
\begin{split}
{\rm X}(t)&=\frac{\Omega_x(t)}{\epsilon_0}(\sin{(2\omega_{RF}t-\phi_x(t))}+\sin{(\phi_x(t))})\\
	  &+\frac{\Omega_y(t)}{\epsilon_0}(-\cos{(2\omega_{RF}t-\phi_y(t))}+\cos{(\phi_y(t))}),
\end{split}
\ee
and
\be
\begin{split}
\Upsilon(t)&=\frac{\Omega_x(t)}{\epsilon_0}(\cos{(2\omega_{RF}t-\phi_x(t))}-\cos{(\phi_x(t))})\\
	  &+\frac{\Omega_y(t)}{\epsilon_0}(\sin{(2\omega_{RF}t-\phi_y(t))}+\sin{(\phi_y(t))}).
\end{split}
\ee
We can thus write
\be
\overline{{\bf f}_1({\bf y},t')}\rightarrow\frac{\bar{\Lambda}}{8}\left[\begin{array}{c} -\bar{F}_y^{(4)}\\ \bar{F}_x^{(4)}\\ 0   \end{array}\right]+\frac{\tilde{\Delta}}{4}\left[\begin{array}{c} -\bar{\rm X}\bar{F}_z^{(4)}\\ -\bar{\Upsilon}\bar{F}_z^{(4)}\\ \bar{\rm X}\bar{F}_x^{(4)}+\bar{\Upsilon}\bar{F}_y^{(4)}   \end{array}\right]
\ee
where
\be
\begin{split}
\bar{\Lambda}&=\frac{1}{2}-\frac{\Omega^2_x(t)}{\epsilon^2_0}\cos{(2\phi_x(t))}-\frac{\Omega^2_y(t)}{\epsilon^2_0}\cos{(2\phi_y(t))}\\
	  &+\frac{\Omega_x(t)\Omega_y(t)}{\epsilon^2_0}\sin{(\phi_x(t)-\phi_y(t))},
\end{split}
\ee
\be
\bar{\rm X}=\frac{\Omega_x(t)}{\epsilon_0}\sin{(\phi_x(t))}+\frac{\Omega_y(t)}{\epsilon_0}\cos{(\phi_y(t))},
\ee
and
\be
\bar{\Upsilon}=-\frac{\Omega_x(t)}{\epsilon_0}\cos{(\phi_x(t))}+\frac{\Omega_y(t)}{\epsilon_0}\sin{(\phi_y(t))}.
\ee

Putting all together, the Heisenberg equations for the components of the total angular momentum, up to second order correction of the RWA, are
\begin{widetext}
\be
\frac{d}{dt'}\left[\begin{array}{c} \bar{F}_x^{(4)}\\ \bar{F}_y^{(4)}\\ \bar{F}_z^{(4)}\end{array}\right]=\frac{\epsilon}{2}\left[\begin{array}{c} \bar{\upsilon}\bar{F}_z^{(4)}+2\Delta_{RF}F_y^{(4)}\\-\bar{\chi}\bar{F}_z^{(4)}-2\Delta_{RF}F_x^{(4)}\\ \bar{\chi}\bar{F}_y^{(4)}-\bar{\upsilon}\bar{F}_x^{(4)}   \end{array}\right]+\epsilon^2\left\{\frac{\bar{\Lambda}}{8}\left[\begin{array}{c} -\bar{F}_y^{(4)}\\ \bar{F}_x^{(4)}\\ 0   \end{array}\right]+\frac{\tilde{\Delta}}{4}\left[\begin{array}{c} -\bar{\rm X}\bar{F}_z^{(4)}\\ -\bar{\Upsilon}\bar{F}_z^{(4)}\\ \bar{\rm X}\bar{F}_x^{(4)}+\bar{\Upsilon}\bar{F}_y^{(4)}   \end{array}\right]\right\}.
\label{eq:RWAsecondorder}
\ee
\end{widetext}
Equivalently, using Eq. (\ref{eq:RWAsecondorder}), we can write the Hamiltonian, Eq. (\ref{eq:H4}), up to second order correction
\begin{widetext}
\be
\begin{split}
H^{(4)}(t)&\approx\left(\frac{3\Omega_0}{2}(1-g_r)+\frac{1}{2}(7\Delta_{RF}-\Delta_{\mu w})\right)I^{(4)}\\
&+\left(\frac{\Omega_x(t)}{2}\left(\cos{(\phi_x(t))}+\frac{\Delta_{RF}}{2\omega_{RF}}\sin{(\phi_x(t))}\right)+\frac{\Omega_y(t)}{2}\left(\sin{(\phi_y(t))}+\frac{\Delta_{RF}}{2\omega_{RF}}\cos{(\phi_y(t))}\right)\right)F_x^{(4)}\\
			   &+\left(-\frac{\Omega_x(t)}{2}\left(\sin{(\phi_x(t))}+\frac{\Delta_{RF}}{2\omega_{RF}}\cos{(\phi_x(t))}\right)+\frac{\Omega_y(t)}{2}\left(\cos{(\phi_y(t))}+\frac{\Delta_{RF}}{2\omega_{RF}}\sin{(\phi_y(t))}\right)\right)F_y^{(4)}\\
			   &+\frac{1}{16\omega_{RF}}\bigg(\Omega_x^2(t)\Big(1-2\cos{(2\phi_x(t))}\Big)+\Omega_y^2(t)\Big(1-2\cos{(2\phi_y(t))}\Big)+2\Omega_x(t)\Omega_y(t)\sin{(\phi_x(t)-\phi_y(t))}\bigg)F_z^{(4)}.
\end{split}
\ee
\end{widetext}
Using a similar procedure to the one detailed above, a second order correction for the Hamiltonian acting on the $F=3$ manifold can also be obtained. Putting all the second order correction terms together, the RF control Hamiltonian for the full ground manifold, in the RWA, corrected up to second order can be written
\begin{widetext}
\be
\begin{split}
H_{RF}(t)&=
\frac{\Omega_x(t)}{2}\left[\cos{(\phi_x(t))}\left(F_x^{(4)}-g_r\left(1-\frac{\Omega_0(1-g_r)}{2\omega_{RF}}\right)F_x^{(3)}\right)-\sin{(\phi_x(t))}\left(F_y^{(4)}+g_r\left(1+\frac{\Omega_0(1-g_r)}{2\omega_{RF}}\right)F_y^{(3)}\right)\right]\\ 
&+\frac{\Omega_x(t)}{2}\left[\frac{\Delta_{RF}}{2\omega_{RF}}\left(\sin{(\phi_x(t))}F_x^{(4)}-g_r\cos{(\phi_x(t))}F_x^{(3)}\right)-\frac{\Delta_{RF}}{2\omega_{RF}}\left(\cos{(\phi_x(t))}F_y^{(4)}+g_r\sin{(\phi_x(t))}F_y^{(3)}\right)\right]\\
&+\frac{\Omega_y(t)}{2}\left[\cos{(\phi_y(t))}\left(F_y^{(4)}-g_r\left(1-\frac{\Omega_0(1-g_r)}{2\omega_{RF}}\right)F_y^{(3)}\right)+\sin{(\phi_y(t))}\left(F_x^{(4)}+g_r\left(1+\frac{\Omega_0(1-g_r)}{2\omega_{RF}}\right)F_x^{(3)}\right)\right]\\ 
&+\frac{\Omega_y(t)}{2}\left[\frac{\Delta_{RF}}{2\omega_{RF}}\left(\cos{(\phi_y(t))}F_x^{(4)}+g_r\sin{(\phi_y(t))}F_x^{(3)}\right)+\frac{\Delta_{RF}}{2\omega_{RF}}\left(\sin{(\phi_y(t))}F_y^{(4)}+g_r\cos{(\phi_y(t))}F_y^{(3)}\right)\right]\\
&+\frac{1}{16\omega_{RF}}\bigg(\Omega_x^2(t)\Big(1-2\cos{(2\phi_x(t))}\Big)+\Omega_y^2(t)\Big(1-2\cos{(2\phi_y(t))}\Big)+2\Omega_x(t)\Omega_y(t)\sin{(\phi_x(t)-\phi_y(t))}\bigg)F_z^{(4)}\\
			   &-\frac{g_r^2}{16\omega_{RF}}\bigg(\Omega_x^2(t)\Big(1-2\cos{(2\phi_x(t))}\Big)+\Omega_y^2(t)\Big(1-2\cos{(2\phi_y(t))}\Big)-2\Omega_x(t)\Omega_y(t)\sin{(\phi_x(t)-\phi_y(t))}\bigg)F_z^{(3)}.
\end{split}
\label{eq:Hrfcomplete}
\ee
\end{widetext}


\end{document}